\documentclass[aps,apl,10]{emulateapj}
\slugcomment{{\sc Accepted to ApJ:} Mar 27, 2012} 

\usepackage{epsfig}
\usepackage{amsmath}
\usepackage{natbib}
\begin{document}

\title{The EVIL-MC Model for Ellipsoidal Variations of Planet-Hosting Stars and Applications to the HAT-P-7 System}

\author{Brian K. Jackson\altaffilmark{1}}
\affil{Carnegie Institution for Science, Washington, DC 20015, USA}
\email{bjackson@dtm.ciw.edu}
\author{Nikole K. Lewis}
\affil{Lunar and Planetary Laboratory, University of Arizona, Kuiper Space Sciences Building, Tucson AZ 85721, USA}
\author{Jason W. Barnes}
\affil{Department of Physics, University of Idaho, Engineering-Physics Building, Moscow, ID 83844, USA}
\author{L. Drake Deming}
\affil{Department of Astronomy, University of Maryland, College Park, Maryland 20742, USA}
\author{Adam P. Showman}
\affil{Lunar and Planetary Laboratory, University of Arizona, Kuiper Space Sciences Building, Tucson AZ 85721, USA}
\and
\author{Jonathan J. Fortney}
\affil{Department of Astronomy and Astrophysics, UCO/Lick Observatory, University of California, Santa Cruz, CA 95064, USA}

\altaffiltext{1}{Carnegie DTM Astronomy Fellow}

\begin{abstract}
We present a new model for \textbf{E}llipsoidal \textbf{V}ariations \textbf{I}nduced by a \textbf{L}ow-\textbf{M}ass \textbf{C}ompanion, the EVIL-MC model\footnote{An IDL version of the model is publicly available at \href{http://www.lpl.arizona.edu/~bjackson/idl_code/index.html}{http://www.lpl.arizona.edu/$\sim$bjackson/idl\_code/index.html}.}. We employ several approximations appropriate for planetary systems to substantially increase the computational efficiency of our model relative to more general ellipsoidal variation models and improve upon the accuracy of simpler models. This new approach gives us a unique ability to rapidly and accurately determine planetary system parameters. We use the EVIL-MC model to analyze Kepler Quarter 0-2 (Q0-2) observations of the HAT-P-7 system, an F-type star orbited by a $\sim$ Jupiter-mass companion. Our analysis corroborates previous estimates of the planet-star mass ratio $q = (1.10 \pm 0.06)\times10^{-3}$, and we have revised the planet's dayside brightness temperature to $2680^{+10}_{-20}$ K. We also find a large difference between the day- and nightside planetary flux, with little nightside emission. Preliminary dynamical+radiative modeling of the atmosphere indicates this result is qualitatively consistent with high altitude absorption of stellar heating. Similar analyses of Kepler and CoRoT photometry of other planets using EVIL-MC will play a key role in providing constraints on the properties of many extrasolar systems, especially given the limited resources for follow-up and characterization of these systems. However, as we highlight, there are important degeneracies between the contributions from ellipsoidal variations and planetary emission and reflection. Consequently, for many of the hottest and brightest Kepler and CoRoT planets, accurate estimates of the planetary emission and reflection, diagnostic of atmospheric heat budgets, will require accurate modeling of the photometric contribution from the stellar ellipsoidal variation. 
\end{abstract}

\keywords{Planets and satellites: fundamental parameters -- Planets and satellites: individual: HAT-P-7}

\section{Introduction}
\indent The Kepler and CoRoT missions have begun a new chapter in time-domain astronomy. Among other results, the phenomenal photometric stabilities, long observational baselines, and high duty cycles of these missions will provide a vast harvest of new exoplanets. Already, the Kepler mission has found 25 planets that have been confirmed and an additional 1,235 planetary candidates \citep{2011ApJ...736...19B}. The stability of the Kepler and CoRoT photometry also allows access to astrophysical signals with amplitudes too small to have been detected previously. 

\indent Too small to have been observed before, the photometric signal of tidal distortion of a star by a close-in planet can now be measured using Kepler and CoRoT data. This signal is usually referred to as an ellipsoidal variation since a tidally distorted body takes an approximately ellipsoidal shape. \citet{2003ApJ...589.1020D} and \citet{2003ApJ...588L.117L} initially suggested that the Kepler mission might observe ellipsoidal variations for many of its targets, and the latter study estimated amplitudes as large as 100 parts per million (ppm) for very short-period hot Jupiters. 

\indent The amplitude of the ellipsoidal variation depends on several key system parameters, including the ratio of the stellar radius to the orbital semi-major axis $a$, the planet-star mass ratio $q$, and the sine of the orbital inclination $\sin i$. If a planet transits its host star, the transit light curve allows accurate determination of $a$ and $\sin i$. If ellipsoidal variations can also be measured for the system, the mass ratio itself can be estimated. Hence, with an estimate of the stellar mass, ellipsoidal variations can help confirm the planetary nature of a transiting companion \citep{2011AJ....142..195S}.

\indent Kepler and CoRoT observations can also provide constraints on the planetary emission and reflection, which can elucidate a planet's atmospheric properties. However, in visible wavelengths monitored by the missions, the contrast between light emitted from a hot close-in planet and a host star is much smaller than it is in the infrared. Consequently, determination of a close-in planet's emitted and reflected flux requires accounting for the stellar ellipsoidal variation. 

\indent Analysis of ellipsoidal variations and eclipses has a long history for close binary stars, where it provides a wealth of information regarding stellar masses, luminosities, and internal structures, among other properties \citep{1959cbs..book.....K}. The effects of tides in such systems (tidal distortions, thermal perturbations, etc.) can be dramatic, but generations of binary star astronomers were hampered by limited, semi-analytic models. The computational power required for highly accurate numerical models was only developed in the last few decades (e.g.\ \citealp{1994PASP..106..921W}). However, for planetary systems, tidal effects are much smaller, owing to the small planet-star mass ratio ($q \leq 0.01$), and so they give rise to much smaller (and possibly less complex) ellipsoidal variations. Consequently, a model for ellipsoidal variations in planetary systems can be greatly simplified relative to more general models appropriate for binary stars.

\indent In this paper, we present a new model for ellipsoidal variations in planetary systems -- the \textbf{E}llipsoidal \textbf{V}ariations \textbf{I}nduced by a \textbf{L}ow-\textbf{M}ass \textbf{C}ompanion (EVIL-MC) model. We incorporate many approximations appropriate for planetary systems, which allow our model to be computationally efficient. Our model uses an alternative approach to other recently applied or developed models. \citet{2010ApJ...713L.145W} discovered ellipsoidal variations in Kepler observations of the HAT-P-7 system and analyzed them using the binary star ELC code \citep{2000AA...364..265O}. That code is state-of-the-art but requires considerable computational resources to model the very small planet-induced ellipsoidal variation. \citet{2010AA...521L..59M} proposed a semi-analytic model involving a Fourier expansion of photometric signals induced by the presence of a planet, including the ellipsoidal variation. \citet{2011AJ....142..195S} and \citet{2011arXiv1110.3512M} applied that model to photometric variations observed for the KOI-13.01 system. The simplicity of this model allows rapid analysis of data, but the relationships between the Fourier coefficients and the system parameters are not all accurately determined, limiting the ability of this model to determine the parameters. Specifically, in their analysis, \citet{2011AJ....142..195S} did not report a planet-star mass ratio based on the ellipsoidal variation. Determining that relationship requires a model that accounts in detail for tidal effects.

\indent For this paper, we tailor our model to Kepler observations of the HAT-P-7 system and constrain the planet's mass, phase function, whence we derive constraints on the atmospheric brightness temperatures. We also highlight the importance of considering the ellipsoidal variations when modeling planetary emission and reflection in visible wavelengths. In Section \ref{sec:mod_desc}, we describe our model, derive the relevant equations, and compare our model to others. In Section \ref{sec:observations}, we describe the Kepler observations of the HAT-P-7 system and how we conditioned the data for analysis. In Section \ref{sec:analysis}, we apply the EVIL-MC model to these data. Finally, in Section \ref{sec:disc}, we discuss implications of our results and future work.

\section{Model Description}
\label{sec:mod_desc}
\indent In this section, we first describe the approximations made in our model and derive the relevant equations. Then, we compare our model to others.

\subsection{Model Approximations}
\label{sec:mod_approx}
To model tidal distortion of stars hosting close-in planets, we make several approximations:
\begin{enumerate}
\item We treat the planet and star as point masses to determine their gravitational fields, ignoring the contribution of the asymmetric mass distributions arising from tidal distortions, which is negligible for our purposes \citep{2000AA...359..289C}. 
\item We employ the equilibrium tide approximation for modeling the stellar shape, in which the stellar surface lies along a gravitational isopotential \citep{1976ApJ...203..182W}. We neglect tidal dissipation or more complex hydrodynamic motions within the star that may be observable in some cases \citep{2008ApJ...679..783P}. 
\item We assume that the planet's orbit is circular. Although the orbits of many transiting planets are eccentric (notably HD 80606 b -- \citealp{2009ApJ...703.2091W}), the majority are circular or nearly so. 
\item We assume that the planet-star mass ratio $q$ is small and that the host star rotates as a solid body with a centrifugal acceleration that is small compared to the surface gravity. As a consequence, the departure of the stellar shape from sphericity due to tides and rotation is assumed small (tens of ppm for the HAT-P-7 system or $\sim$ 10 km).
\item We neglect Doppler effects from the velocity of tidal motions within the star. The rotational and tidal motions of the stellar surface may contribute to the Doppler flux variations \citep{2012MNRAS.tmp.2682A} but are probably negligible for broadband Kepler observations.
\end{enumerate}

\indent We do NOT assume that the stellar rotation is synchronous with the orbit or that the stellar obliquity is zero. Alignment and synchronization of stellar rotation with the orbit is common among close binary stars, likely as a result of large tidal torques \citep{1959cbs..book.....K}, but not for transiting planets. For the majority of planet-hosting stars for which it can be determined, the stellar rotation period is much longer than the planet's orbital period \citep{2008ApJ...678.1396J}. Also, observations of the Rossiter-McLaughlin effect \citep{2010ApJ...718L.145W}, detection of the crossing of star spots by transiting planets (e.g.\ \citealp{2011ApJ...740...33D}), and contributions of gravity darkening to transit light curves (e.g.\ \citealp{2011ApJS..197...10B}) all show that many planets have orbits that are strongly inclined relative to their host stars' equators.

\indent We also do NOT make the usual assumption in tidal modeling that the orbital separation is much larger than the physical radius of the star, i.e.\ we do NOT assume the star is an ellipsoid. Typically, the potential of a body inducing tidal distortion (and consequently the shape of the tidally distorted body) is expanded in the ratio of the tidally-distorted body's radius to the orbital separation $a$ \citep{1999ssd..book.....M}. For planets very close to their host star -- the planets for which ellipsoidal variation will be the most pronounced -- higher-order terms may contribute to the distortion non-negligibly. For the HAT-P-7 system, $a \sim 4$ \citep{2008ApJ...680.1450P}, so departure of the stellar shape from an ellipsoid is significant. \citet{2008ApJ...679..783P} showed that assuming the star is an ellipsoid may not be sufficiently accurate for very close-in exoplanets, resulting in erroneous estimates of the system parameters. We discuss this point more in Section \ref{sec:comp_models}.

\indent Currently our model does NOT include occultation of either the star (during transit) or the planet (during eclipse). In fitting our model to data, we mask out these phases. A future version of our model will include these phases, but an initial analysis suggests that, for typical planetary systems, the tidal distortion of the host star negligibly modifies the transit light curve. Thus, standard light curve models (e.g.\ \citealp{2002ApJ...580L.171M}) are probably sufficiently accurate.

\subsection{Model Equations}
\begin{figure}
\includegraphics{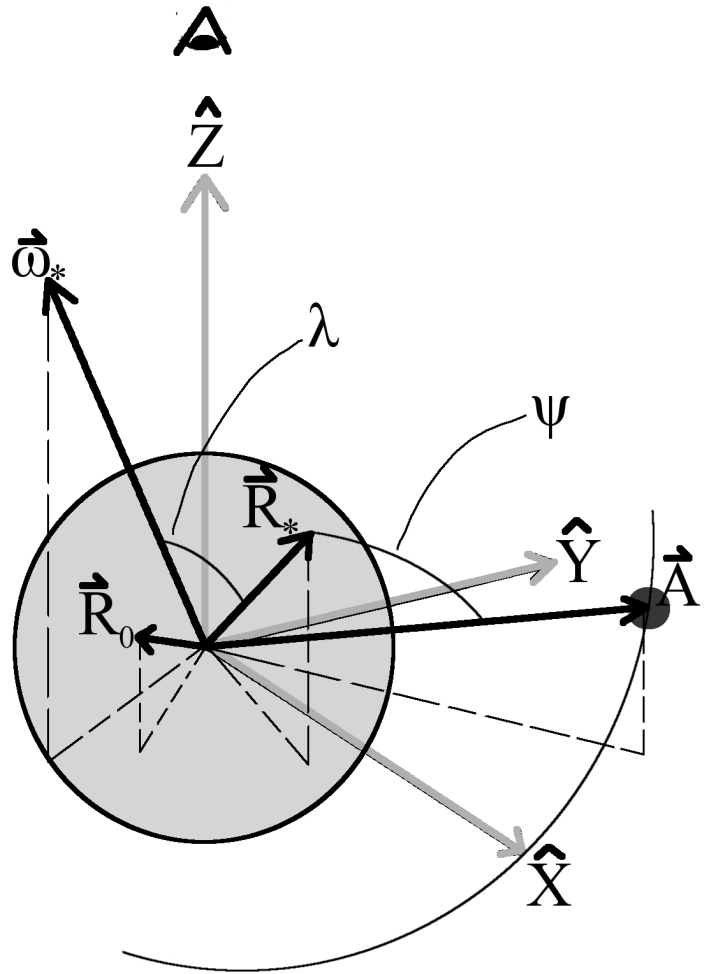}
\caption{Definition of model geometry. The large, light circle represents the star, and the small, dark circle represents the planet. The coordinate system ($X$, $Y$, $Z$) is centered on the star, $\mathbf{\hat{Z}}$ points toward the observer, $\mathbf{\hat{X}}$ points along the orbital line of nodes, and $\mathbf{\hat{Y}}$ is in the plane of the sky and perpendicular to $\mathbf{\hat{X}}$. $\mathbf{R}_\star$ points somewhere on the stellar surface, $\mathbf{A}$ points to the planet in its orbit (a portion of which is illustrated), and $\mathbf{\omega}_\star$ is the stellar rotation vector. $\mathbf{R}_0$ points somewhere on a stellar surface at a right angle to both $\mathbf{\omega}_\star$ and $\mathbf{A}$. The relevant angles are $\psi$, the angle between $\mathbf{R}_\star$ and $\mathbf{A}$, and $\lambda$, the angle between $\mathbf{R}_\star$ and $\mathbf{\omega}_\star$.}
\label{fig:prob_geom}
\end{figure}

\indent We use a coordinate system centered on the star, as illustrated in Figure \ref{fig:prob_geom}. (In our notation, a vector $\mathbf{Q}$ has magnitude $Q$ and is parallel to $\mathbf{\hat{Q}}$, a vector with unit length.) As measured in this frame (and subject to the approximations above), the gravitational potential on the stellar surface $U$ is:
\begin{eqnarray}
U = \frac{G M_\star}{R_\star} + \frac{G M_p}{\left(A^2-2R_\star A \cos \psi + R_\star^2\right)^{1/2}} \nonumber \\ - \frac{G M_p}{A^2}R_\star \cos \psi + \frac{1}{2} \omega_\star^2 R_\star^2 \left(1 - \cos^2\lambda\right)
\label{eqn:roche_pot_rot_unred}
\end{eqnarray}
where $G$ is Newton's gravitational constant, $M_\star$ the stellar mass, $R_\star$ the distance from the stellar center to its photosphere (which is not constant), $M_p$ the planet's mass, $A$ the orbital semi-major axis, $\mathbf{\omega_\star}$ the stellar rotation rate,  $\cos \psi = \mathbf{\hat{R}_\star} \cdot \mathbf{\hat{A}}$, and $\cos \lambda = \mathbf{\hat{R}_\star} \cdot \mathbf{\hat{\omega}_\star}$. The stellar rotation axis is $\mathbf{\omega_\star}$, and the planet's position vector is $\mathbf{A}$.

\indent The first term in Equation \ref{eqn:roche_pot_rot_unred} represents the star's gravitational potential, and the second term the planet's gravitational potential. If the second term in Equation \ref{eqn:roche_pot_rot_unred} were expanded as a Taylor series in $R_\star/A$, the first-order term (proportional to $\cos \psi$) would correspond to the force constant throughout the star that keeps it in orbit about the system barycenter. Since this force is constant, it does not contribute to the tidal distortion, and so we include the third term in Equation \ref{eqn:roche_pot_rot_unred} to remove it. The fourth term in Equation \ref{eqn:roche_pot_rot_unred} represents a potential corresponding to the centrifugal acceleration due to the stellar rotation. 

\indent At points on the stellar surface where $\mathbf{R_\star} \perp \mathbf{A}$ and $\mathbf{R_\star} \perp \mathbf{\omega_\star}$, we take $\mathbf{R_\star} \equiv \mathbf{R_0}$. To clarify this definition, consider the case of a planet crossing the exact center of the disk of a star with a rotation vector pointing exactly at the observer. At this instant, $\mathbf{\hat{A}} \parallel \mathbf{\hat{\omega}_\star} \parallel \mathbf{\hat{Z}}$, and $R_0$ would be the usual stellar radius that goes into determining the transit depth. In the general case, the relationship between the transit depth and $R_0$ is more complicated. However, the tidal distortion has a negligible effect on the transit light curve, so we can safely consider $R_0$ as the usual radius that goes into determining the transit depth. 

\indent We normalize $U$ by $\left(\frac{G M_\star}{R_0}\right)$, giving $\Phi$: 
\begin{eqnarray}
\Phi \equiv U \left(\frac{R_0}{G M_\star}\right) = \frac{1}{R} + \frac{q}{\left(a^2-2 a R \cos \psi + R^2\right)^{1/2}} \nonumber \\ - \frac{q}{a^2} R \cos \psi + \frac{1}{2} \frac{\omega^2 R^2}{a^3}\left(1+q\right)\left(1-\cos^2\lambda\right)
\label{eqn:roche_pot_rot}
\end{eqnarray}
where $R = R_\star/R_0$, $q = M_p/M_\star$, $a = A/R_0$, and $\omega = \omega_\star/n$, with $n$ as the orbital mean motion. 

\indent Normalized to $R_0$, the stellar radius $R = 1 + \delta R$, a function of $\cos \psi$ and $\cos \lambda$. Per our assumptions, the surface of the star corresponds to an isopotential contour, i.e.\ $\Phi =$ const. We take the constant to be the potential $\Phi_0$ at $\mathbf{R_0}$ (where $\cos \psi = \cos \lambda = 0$):
\begin{equation}
\Phi_0 = 1 + \frac{q}{\left(a^2 + 1\right)^{1/2}} + \frac{1}{2}\frac{\omega^2}{a^3} (1+q).
\label{eqn:Phi_0}
\end{equation}
The departure from sphericity $\delta$R is assumed small, and we can expand $\Phi$: 
\begin{align}
\Phi &= \frac{1}{1 + \delta R} + \frac{q}{\left(a^2-2 a (1 + \delta R) \cos \psi + (1 + \delta R)^2\right)^{1/2}} \nonumber \\ &- \frac{q}{a^2} (1 + \delta R) \cos \psi + \frac{1}{2} \frac{\omega^2 (1 + \delta R)^2}{a^3}\left(1+q\right)\left(1-\cos^2\lambda\right) 
\nonumber \\  &\approx \left(1 - \delta R\right) + \frac{q}{\left(a^2-2 a \cos \psi + 1\right)^{1/2}} - \frac{q}{a^2} \cos \psi \nonumber \\  &+ \frac{1}{2} \frac{\omega^2}{a^3}\left(1-\cos^2\lambda\right)           
\label{eqn:expand_phi}
\end{align}
and $\Phi_0$:
\begin{equation}
\Phi_0 \approx 1 + \frac{q}{\left(a^2 + 1\right)^{1/2}} + \frac{1}{2}\frac{\omega^2}{a^3}.
\label{eqn:expand_Phi0}
\end{equation}

\indent In Equations \ref{eqn:expand_phi} and \ref{eqn:expand_Phi0}, we have dropped 2nd-order terms. We set Equation \ref{eqn:expand_phi} equal to \ref{eqn:expand_Phi0} and solve for $\delta R$: 
\begin{align}
\delta R = q &\left([a^2 - 2 a \cos \psi + 1]^{-1/2} - [a^2 + 1]^{-1/2} - \frac{\cos \psi}{a^2}\right) \nonumber \\ &- \frac{\omega^2}{2a^3} \cos^2\lambda.
\label{eqn:del_R}
\end{align}

\indent Gravity darkening of the stellar surface also contributes to the photometric variation. Briefly, the planet's tidal gravity perturbs the balance of forces (pressure, the star's own gravity, radiation, etc.) within the stellar atmosphere and results in a small (few 0.1 K) decrease in the effective temperature and brightness at points on the stellar surface nearest the planet \citep{1924MNRAS..84..665V}. Theoretical considerations motivate a parameterization of the gravity darkening  involving the surface gravity \citep{1971ApJ...166..605W}. The gravity vector on the stellar surface is given by 
\begin{align}
\mathbf{g} = &-\frac{G M_\star}{R_\star^2}\mathbf{\hat{R}_\star} + \frac{G M_p \left(\mathbf{A}-\mathbf{R_\star}\right)}{\left(A^2-2R_\star A\cos \psi + R_\star^2\right)^{3/2}} - \frac{G M_p}{A^3}\mathbf{A} \nonumber \\ &+ \omega_\star^2 R_\star\left(\mathbf{\hat{R}_\star}-\mathbf{\hat{\omega}_\star} \cos \lambda\right).
\label{eqn:grav_unred}
\end{align}
The last term in Equation \ref{eqn:grav_unred}, representing the centrifugal acceleration, is usually written as $-\mathbf{\omega_\star} \times (\mathbf{\omega_\star} \times \mathbf{R_\star})$. Using cross-product identities, this expression can be re-written as
\begin{equation}
\label{eqn:cross_prod_iden}
\mathbf{R}_\star (\mathbf{\omega}_\star \cdot \mathbf{\omega}_\star) - \mathbf{\omega}_\star (\mathbf{\omega}_\star \cdot \mathbf{R}_\star) = \nonumber \\
\omega_\star^2 R_\star (\mathbf{\hat{R}_\star} - \mathbf{\hat{\omega}_\star} \cos \lambda).
\end{equation}

\indent We normalize $\mathbf{g}$ by $\left(\frac{G M_\star}{R_0^2}\right)$, giving $\mathbf{\Gamma}$:
\begin{align}
\mathbf{\Gamma} \equiv \mathbf{g} \left(\frac{R_0^2}{G M_\star}\right) &= -\frac{1}{R^2} \mathbf{\hat{R}_\star} + \frac{q \left(a \mathbf{\hat{A}} - R \mathbf{\hat{R}_\star}\right) }{\left(a^2 - 2a R \cos \psi + R^2\right)^{3/2}} \nonumber \\ &- \frac{q}{a^2} \mathbf{\hat{A}} + \omega^2 \frac{R}{a^3} \left(1 + q\right) \left(\mathbf{\hat{R}_\star}-\mathbf{\hat{\omega}_\star} \cos \lambda\right)
\nonumber \\
&\approx -(1 - 2\, \delta R) \mathbf{\hat{R}_\star} + \frac{q \left(a \mathbf{\hat{A}} - \mathbf{\hat{R}_\star}\right)}{\left(a^2 - 2a \cos \psi + 1\right)^{3/2}}  \nonumber \\ &- \frac{q}{a^2} \mathbf{\hat{A}} + \frac{\omega^2}{a^3} \left(\mathbf{\hat{R}_\star}-\mathbf{\hat{\omega}_\star} \cos \lambda\right) 
\nonumber \\
&\equiv -\mathbf{\hat{R}_\star} + \mathbf{\delta \Gamma}
\label{eqn:grav}
\end{align}
where $\mathbf{\delta \Gamma}$ represents all the gravitational accelerations other than the zeroth-order stellar gravity. The magnitude of $\mathbf{\Gamma}$ is
\begin{align}
\Gamma &= \left(\mathbf{\Gamma} \cdot \mathbf{\Gamma}\right)^{1/2} 
\nonumber\\
&= \left([-\mathbf{\hat{R}_\star} + \mathbf{\delta \Gamma}] \cdot [-\mathbf{\hat{R}_\star} + \mathbf{\delta \Gamma}]\right)^{1/2} 
\nonumber \\
&\approx 1 - \mathbf{\hat{R}_\star} \cdot \mathbf{\delta \Gamma}.
\label{eqn:mag_gam}
\end{align}

\indent Likewise, at $\mathbf{R}_0$, the gravity vector is $\Gamma_0 \approx 1 - \mathbf{\hat{R}_0} \cdot \mathbf{\delta \Gamma_0}$. Using these expressions, the effective temperature at $T$ on the stellar surface is parameterized as
\begin{equation}
T = T_\star \left(\frac{\Gamma}{\Gamma_0}\right)^\beta \simeq T_\star \left(1 + \beta [\mathbf{\hat{R}_0} \cdot \mathbf{\delta \Gamma}_0 - \mathbf{\hat{R}}_\star \cdot \mathbf{\delta \Gamma}]\right)
\label{eqn:Teff}
\end{equation}
where $T$ is the temperature at $\mathbf{R}_\star$, and $\beta$ the gravity darkening exponent. $T_\star$ is the effective temperature at $\mathbf{R}_0$. For our analysis, we are only interested in the fractional variation in the stellar brightness, and surface brightness variations are linear in the small difference in temperature between $\mathbf{R}_0$ and any other point on the surface. Moreover, tides raised by planets have a negligible effect on the determination of the stellar effective temperature from observation, and the usual distinctions between a star's polar and mean effective temperatures \citep{1979ApJ...234.1054W} are unimportant here. Consequently, we take $T_\star$ to be both the mean effective temperature and the temperature at $\mathbf{R}_0$.

\indent To model the limb-darkening of the stellar disk, we calculate the projection of the normalized gravity vector onto the line of sight, $\mu = \mathbf{\hat{\Gamma}} \cdot \mathbf{\hat{Z}}$. We use this to determine the limb-darkening profile $I(\mu)$ assuming a quadratic profile \citep{2002ApJ...580L.171M}:
\begin{equation}
I(\mu)/I(1) = 1 - \gamma_1 (1 - \mu) - \gamma_2  (1 - \mu)^2
\label{eqn:LDC_prof_quad}
\end{equation}
where $\gamma_i$ are the limb-darkening coefficients. The model DOES allow for other profiles, though.

\indent Our model also includes the photometric effects of the stellar reflex velocity $v_Z$, referred to as ``Doppler flux variations'' in \citet{2003ApJ...588L.117L}. These variations come in at the first order in the ratio of $v_Z$ to the speed of light and include several effects convolved together: (1) transformation of the energy-momentum four-vector from the frame co-moving with the star to the observer's frame (Equation 4.93 from \citealt{1979rpa..book.....R}), (2) reduction in the apparent size of the star as it recedes from the observer (Equation 4.95 from \emph{ibid.}), (3) increased travel time for the stellar photons as the star recedes from the observer (see discussion point 2 above Equation 4.97 in \emph{ibid.}), and (4) Doppler shifting of the stellar flux measured within the observational bandpass. Together, effects (1)-(3) increase the apparent stellar flux as the star approaches the observer. The peak in emission for HAT-P-7 occurs blueward of the Kepler bandpass, so the accompanying blue-shift of the stellar flux (effect 4) reduces the apparent flux. However, taken altogether, the Doppler flux variations cause HAT-P-7 to brighten as it approaches and darken as it recedes.

\indent To first order in $q$, the star's line-of-sight velocity is
\begin{align}
v_Z &= -(q \sin i) n A \sin(2 \pi \phi) \nonumber \\ &= -\left(\frac{2\pi G M_\star}{P}\right)^{1/3} (q \sin i) \sin(2 \pi \phi) \nonumber \\ &= -K_Z \sin(2 \pi \phi)
\label{eqn:reflex_vel}
\end{align}
where $A$ is the orbital semi-major axis (NOT normalized to $R_0$), $\phi$ is the orbital phase (= 0 at mid-transit), $P$ is the orbital period, and $K_Z$ is the amplitude of the projected reflex velocity of the star. (Note that we have chosen the opposite sign convention from \citealt{2003ApJ...588L.117L}: positive $v_Z$ corresponds to increasing radial distance.) 

\indent For our model, we tile the stellar surface in lat/long. To first order in $\delta R$, each tile's projected area $\Delta A_p$ is 
\begin{equation}
\Delta A_p = \left(1 + 2\delta R\right) \mu \cdot \Delta \Omega
\label{eqn:proj_area}
\end{equation} 
where $\Delta \Omega$ is the solid angle of each grid point.

\begin{figure*}
\centering
\includegraphics[width=0.8\textwidth]{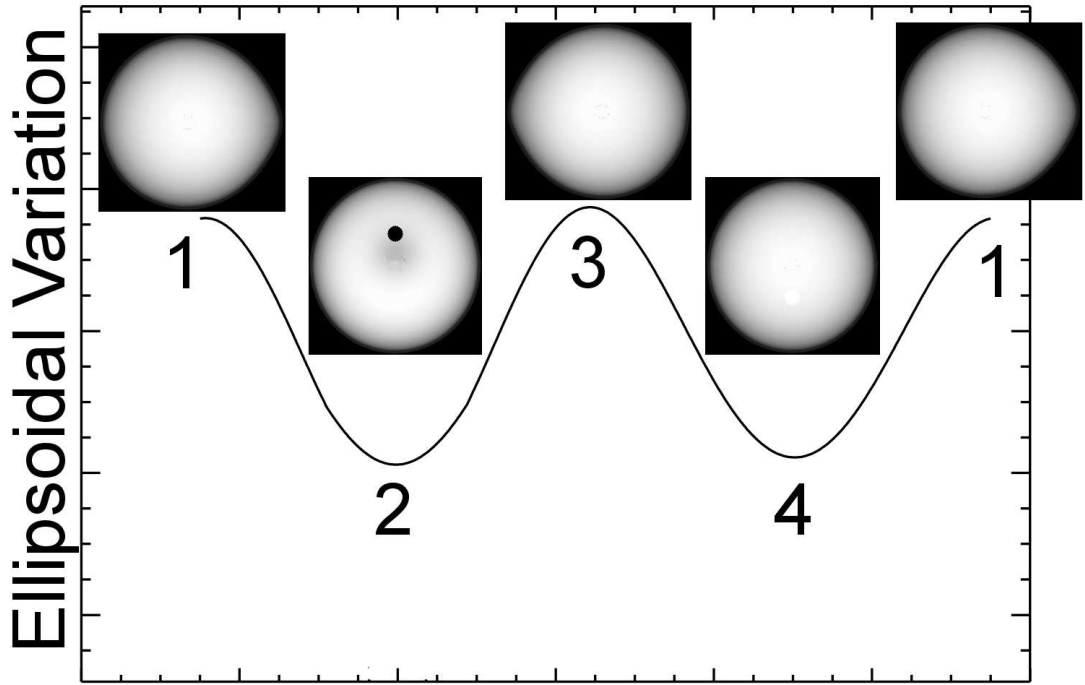}
\caption{Cartoon illustrating the phases of ellipsoidal variation. The y-axis is in arbitrary units, and the x-axis is orbital phase. The planet-star mass ratio is exaggerated for illustrative purposes. Phases 1 ($\phi = -0.25$) and 3 ($\phi = 0.25$) show the planet at quadrature, phase 2 ($\phi = 0$) shows the planet during transit (the photometric signature of which is NOT included), and phase 4 ($\phi = 0.75$) shows the planetary eclipse.}
\label{fig:cartoon}
\end{figure*}

\indent For a given orbital phase $\phi$, we calculate $\delta$R, $\delta \Gamma$, and $T$ at each grid point on the stellar hemisphere visible to the observer ($Z \ge 0$), along with $\mathbf{\hat{R}_0} \cdot \mathbf{\delta \Gamma_0}$. To include the Doppler flux variation, we assume each point on the star is a blackbody -- \citet{2003ApJ...588L.117L} showed that departure from blackbody emission changes the photometric signature of the ellipsoidal variation by only a few percent. We use Equation 2 from \citet{2003ApJ...588L.117L} to calculate the monochromatic flux throughout the Kepler observational bandpass (\href{http://keplergo.arc.nasa.gov/CalibrationResponse.shtml}{http://keplergo.arc.nasa.gov/CalibrationResponse.shtml}) and convolve the flux with the Kepler response function. (Note that the Kepler response function is given in wavelength space, so we had to convert it to frequency space to use the results from \citealt{2003ApJ...588L.117L}.) Using the emission calculated for each point on the stellar surface, we multiply each point's flux by the appropriate limb-darkening profile value ($I(\mu)/I(1)$) and sum the contributions from all grid points. Finally, we move to the next point in the orbit and do the calculation over again. Figure \ref{fig:cartoon} illustrates schematically the appearance of the distorted star and the resulting photometric variations.

\indent Because our model is linearized in small quantities, it is computationally more efficient for planet-star systems than more general models, particularly those designed for binary stars (e.g.\ \citealp{2000AA...364..265O}). Where more general models require a few hundred thousand grid elements to accurately model the ellipsoidal variation (e.g.\ \citealp{2010ApJ...713L.145W}), our model requires only a few hundred for convergence to better than 0.01 ppm. To check the accuracy of our approximations, we compared the values for all calculated physical quantities (radius, temperature, etc.) as determined by the linearized equations and the exact equations. For the HAT-P-7 system, all quantities converged to better than 10 parts per billion. 

\indent The ellipsoidal variation signal is convolved with the planet's reflection and emission in the Kepler data. Therefore, to fit the Kepler data, we also require a model for light emerging from a planet. In visible wavelengths, the light emerging from a close-in planet is likely dominated by reflection of stellar radiation, which suggests the planetary phase function should be nearly symmetric about $\phi = 0.5$. Although it can be more complicated (e.g.\ \citealp{2008ApJ...678L.129C}), we take a simple sinusoidal phase curve for the planet:
\begin{equation}
F_p = F_0 - F_1 \cos (2 \pi \phi)
\label{eqn:pl_phs_crv}
\end{equation}
where $F_0$ is a constant term, $F_1$ is the amplitude of the planet's reflected and emitted light, and both are ratioed to the stellar emission at mid-eclipse (when the planet is occulted). The sum $F_0 + F_1$ can be directly estimated from the depth of the secondary eclipse.

\indent Close-in planets also suffer significant tidal distortion. This distortion may increase the planet's phase curve at quadrature as a planet's projected surface area is largest there, and this effect may be observable, particularly in the IR \citep{2012ApJ...747...82C}. However, we assume this effect is negligible in Kepler's bandpass. Future work should re-visit this assumption.

\indent The ellipsoidal variation depends on several key system parameters, although there is degeneracy between some parameters ($\beta$ and $q$, for example). Given sufficiently high quality data, the ellipsoidal variation can be used to determine at least seven parameters: $q$, $a$, $\mathbf{\omega}_\star$, $\sin i$, $\gamma_i$, $K_Z$, and $\beta$. A fit to the planet's phase curve determines $F_0$ and $F_1$. We expect the planet's phase curve to oscillate with the orbital period and the ellipsoidal variation to oscillate with half the orbital period (twice an orbit). While they have a period equal to the orbital period, the Doppler flux variations are 90$^\circ$ out of phase with the planet's phase function. Thus, in principle, analysis of Kepler observations should be able to distinguish these different components (given that the planet's phase curve is symmetric about $\phi = 0.5$). 

\indent There is no general expression relating the physical parameters to the amplitude of the ellipsoidal variation, but approximating the star's shape as an ellipsoid, we can explicitly express the ellipsoidal variation's dependence on system parameters. Then the photometric oscillations can be expanded as a Fourier series. Combining equations from \citet{2010AA...521L..59M} and \citet{1985ApJ...295..143M} gives the following series for the combined ellipsoidal variation, Doppler flux variations, and planet's phase curve $\frac{\Delta F}{F}$:
\begin{align}
\frac{\Delta F}{F} = &-A_{ellip} \cos (2 \cdot 2 \pi \phi) + A_{beam} \sin (2 \pi \phi) \nonumber \\ &- A_{refl} \cos (2 \pi \phi)
\label{eqn:fourier_series}
\end{align}
where $A_{ellip} = \alpha_{ellip} (q \sin^2 i) a^{-3}$, $A_{beam} = \alpha_{beam} 4 \left(\frac{K_Z}{c}\right)$, and $A_{refl} = p_{geo} \left(\frac{R_p}{A}\right)^2$. Here $p_{geo}$ is the planet's geometric albedo, and $R_p$ the planet's radius. $\alpha_{ellip}$ depends on the gravity-darkening and limb-darkening coefficients, and $\alpha_{beam}$ corrects the amplitude of the Doppler flux variations for shifting of flux into and out of the observational bandpass. Both $\alpha$s are of order unity \citep{2010AA...521L..59M}, but more accurate estimates are required to provide estimates of, for example, $q$. Moreover, as discussed in Section \ref{sec:mod_approx}, the above Fourier series less accurately approximates the photometric oscillation for very close-in planets ($a \rightarrow 1$), as higher-order harmonics contribute non-negligibly. 

\indent Although, in principle, ellipsoidal variations can constrain $a$, $i$, and $\gamma_i$, we expect that transit observations (if a planet DOES transit) will provide tighter constraints. Also, ellipsoidal variations are relatively insensitive to $\beta$, and so modeling based on spectral characterization of a star may provide better estimates (e.g.\ \citealp{2011AA...529A..75C}). On the other hand, when ellipsoidal variations can constrain $q$ and transit observations $\sin i$, the Doppler variation signal or radial velocity observations may provide independent constraints on $M_\star$. Consequent to these considerations, in our analysis of the HAT-P-7 observations (Section \ref{sec:analysis}), we do not fit for $a$, $\mathbf{\omega}_\star$, $i$, $\gamma_i$ or $\beta$ and fix these at values provided by other studies.

\indent  We performed several tests to verify that our model works correctly. For example, in the next section, we compare our model to the more general Wilson-Devinney model \citep{2007ApJ...661.1129V} and find good agreement. We also used the results from \citet{2011AJ....142..195S} to test our model for Doppler flux variations (see Equation \ref{eqn:reflex_vel} and preceding discussion). \citet{2011AJ....142..195S} analyzed Doppler flux variations observed for the KOI-13 system and determined their amplitude to be 9.32 ppm,  corresponding to $K_Z = 954$ m/s (see their Equation 1). Our model indicates that $K_Z = 973$ m/s is required to produce that amplitude for that system, within 2\% of the result from \citet{2011AJ....142..195S}.

%
%
%
%

\subsection{Comparison to Other Models}
\label{sec:comp_models}
\indent In this section, we compare our model to previously developed models. We consider the sinusoidal model proposed by \citet{2010AA...521L..59M} and described by Equation \ref{eqn:fourier_series}. Given the assumptions under which they're derived, we expect the sinusoidal model to be less accurate for $a \rightarrow 1$ and the EVIL-MC model to be less accurate as $q \rightarrow 1$. We also consider the publicly available Wilson-Devinney (W-D) model (ftp://ftp.astro.ufl.edu/pub/wilson/), which has a storied history and has been developed for a wide variety of astrophysical circumstances \citep{2007ApJ...661.1129V}. Comparison to other models would be helpful, but the ELC model \citep{2000AA...364..265O} isn't publicly available. The JKTEBOP model \citep{2004MNRAS.351.1277S} IS available (\href{http://www.astro.keele.ac.uk/jkt/codes/jktebop.html}{http://www.astro.keele.ac.uk/jkt/codes/jktebop.html}) but approximates tidally distorted bodies as ellipsoids and so probably would not provide a more accurate description of tidal distortion than the sinusoidal model does\footnote{After this paper was accepted for publication, Dr. Jan Budaj made us aware of another relevant model described in \citet{2011AJ....141...59B} and available at \href{http://www.ta3.sk/~budaj/shellspec.html}{http://www.ta3.sk/$\sim$budaj/shellspec.html}.}. 

\indent Although the W-D model is widely applicable, for modeling planet-induced ellipsoidal variations, its numerical precision is limited to a few tens of ppm (R. E. Wilson, private communication, 2012). Consequently, in the comparison below, the smallest $q$-value we consider is 0.05, which corresponds, for example, to a 50 Jupiter mass body orbiting a solar-mass star. For the range of relevant $a$-values, $q$-values more appropriate to planets ($q \le 10^{-3}$) produce ellipsoidal variations below the W-D model's numerical precision. In any case, the range of $q$ available is sufficient for our purposes.

\indent For the comparison, we fix several model parameters. Neither the W-D nor the sinusoidal model allow quadratic limb-darkening, so we assume linear limb-darkening for the comparison, with a coefficient $u = 0.551$ \citep{2011AA...529A..75C}. (For EVIL-MC, this assumption is equivalent to $\gamma_1 = 0.551, \gamma_2 = 0$.) We take the gravity-darkening coefficient to be $\beta = 0.071$ (corresponding to $g = 4 \beta = 0.284$ for W-D). With these parameters, $\alpha_{ellip} = 0.15 \left(15 + u\right) \left(1 + g\right)/\left(3 - u\right) = 1.223$. We do not include Doppler flux variations and reflected/emitted light from the planet for this comparison. For the sinusoidal model, this assumption requires $A_{beam} = A_{refl} = 0$. For the W-D model, we set the secondary's luminosity ($L_2$) to zero. We also assume no stellar rotation. Unless specified below, all other system parameters are fixed at the values in Table \ref{tbl:fixed_params}.

%
%

\begin{figure}
\includegraphics[width=0.45\textwidth]{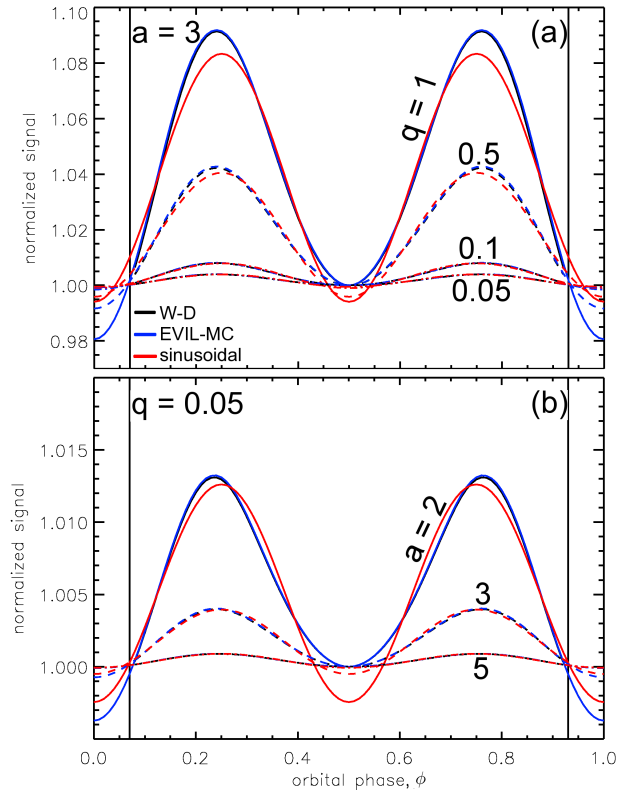}
\caption{Ellipsoidal variations predicted by the Wilson-Devinney (W-D) (black lines), the EVIL-MC (blue), and the sinusoidal (Equation \ref{eqn:fourier_series}) (red) models for a range of $q$ and $a$. The models here do NOT include Doppler flux variations, reflected/emitted light from the secondary (planet), or the transits/eclipses. (The transit phase occurs to the outside of the black, vertical lines.) The models DO include limb- and gravity-darkening. (a) Predictions for fixed $a = 3$ and $q$ equal to 0.05 (dash-dot-dot-dot lines), 0.1, (dash-dot), 0.5 (dash), and 1 (solid). (b) Predictions for fixed $q = 0.05$ and $a$ equal to 2 (solid lines), 3 (dash), and 5 (dash-dot). The difference between the EVIL-MC and W-D models is 2.1\% or less of the total variation and is as good as the agreement can be, given limits on the W-D model's numerical precision. By contrast, the difference between the sinusoidal and W-D models is usually greater than 10\% of the total variation.}
\label{fig:comp_models}
\end{figure}

\begin{table}
\centering
\caption{Model fit parameters from our analysis \label{tbl:fit_params}}
\begin{tabular}{|c|c|c|}
\tableline
\textbf{param.} &\textbf{value (fixed D)} & \textbf{value (var. D)}\\
\tableline
$q$ & $(1.10 \pm 0.06) \times 10^{-3}$ & $(0.99 \pm 0.07) \times10^{-3}$\\
\tableline
$D$ & $61\pm 3$ ppm & $65\pm 2$ ppm\\
\tableline
$T_{day}$ & $2680^{+10}_{-20}$ K & $2700 \pm 10$ K\\
\tableline
$F_1$ & $30 \pm 1$ ppm & $32 \pm 1$ ppm\\
\tableline
$F_0 - F_1$ & 0 or 1 ppm & $3 \pm 3$ ppm\\
\tableline
$K_Z$ & $300 \pm 70$ m/s & $300 \pm 70$ m/s \\
\tableline
\end{tabular}
\tablecomments{The middle column shows best-fit values for fixed $D$ and fixed/variable $K_Z$, while the rightmost column shows values for variable $D$.}
\end{table}

\indent First, we compare results from the three models for a range of $q$ and $a = 3$, as illustrated in Figure \ref{fig:comp_models} (a). To determine the overall normalization for the sinusoidal model, we added an offset value to Equation \ref{eqn:fourier_series} and used a Levenberg-Marquadt (LM) scheme \citep{2009ASPC..411..251M}\footnote{We used Craig Markwardt's mpfit.pro IDL routine, available at \href{http://www.physics.wisc.edu/~craigm/idl/fitting.html}{http://www.physics.wisc.edu/$\sim$craigm/idl/fitting.html}.} to find the value that provided the best agreement between the sinusoidal and W-D models.

\indent As illustrated in Figure \ref{fig:comp_models} (a), agreement between the W-D and EVIL-MC models is better than 2.1\% of the total ellipsoidal variation for all $q$ illustrated, even though the EVIL-MC model is derived under the assumption of small $q$. For $q = 0.05$, the two models agree to 1.1\% of the ellipsoidal variation, corresponding to a difference of about 40 ppm. This discrepancy is near the numerical precision limit of the W-D model and so is as good as the agreement can be. These results indicate the EVIL-MC model is sufficiently accurate to model tidal distortions even in binary systems with stars of comparable mass. By contrast, the sinusoidal model agrees with the W-D model to only about 10\% of the total variation. 

\indent Next, we compare results for a range of $a$ and $q = 0.05$, as illustrated in Figure \ref{fig:comp_models} (b). Agreement between the EVIL-MC and W-D models is better than 1.5\% of the total variation, while agreement between the W-D and sinusoidal model is no better than 8\% and as bad as 20\% (for $a = 2$). As expected, the sinusoidal model is less accurate as $a \rightarrow 1$ as higher order Fourier components contribute more. Whether there exist planets with $a = 2$ and the sinusoidal model can be applied to them remains to be seen (tidal decay of their orbits would probably be rapid -- \citealp{2009ApJ...692L...9L, 2009ApJ...698.1357J}), but the Kepler mission has announced candidates with $a \sim 2$.
 
\begin{figure}
\includegraphics[width=0.45\textwidth]{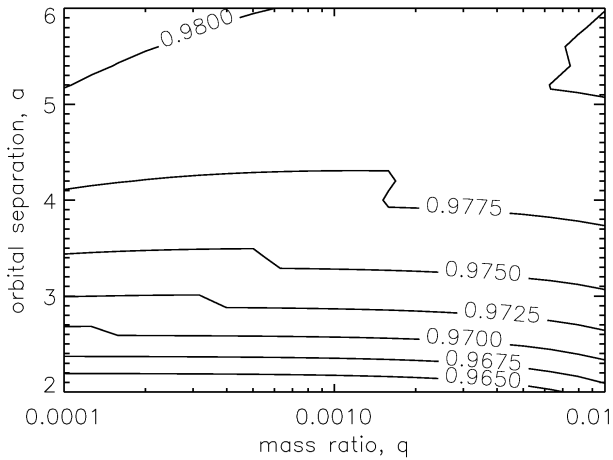}
\caption{Ratio of the $q$-value estimated using the sinusoidal model (Equation \ref{eqn:fourier_series}) and the actual $q$-value (shown along the x-axis) for a range of $q$ and $a$. The sinusoidal model always underestimates the actual mass ratio by a few percent, and the estimate's accuracy degrades for $a \rightarrow 1$ as higher order Fourier components contribute more. }
\label{fig:fit_sine_to_EVILMC}
\end{figure}

\indent We can ask how accurate are estimates of system parameters from the sinusoidal model, particularly the mass ratio. Figure \ref{fig:fit_sine_to_EVILMC} illustrates the accuracy of the $q$-value estimated using the sinusoidal model. For that figure, we calculated ellipsoidal variations for a range of $a$- and $q$-values using the EVIL-MC model (again, neglecting Doppler flux variations or reflected/emitted light from the planet). Then, we used an LM scheme to determine a best-fit $A_{ellip}$ (and offset value) for each modeled ellipsoidal variation and estimated $q$ from $A_{ellip}$ by using the assumed values for all other system parameters ($a$, $\alpha_{ellip}$, etc.). 

\indent Figure \ref{fig:fit_sine_to_EVILMC} shows the ratio of the $q$-value estimated in this way to the actual value. The sinusoidal model typically underestimates $q$ by a few percent, and, as expected, the estimates become less accurate for small $a$. Although estimates of $q$ from, for example, Kepler data are likely to be less accurate than a few percent, estimates of $q$ using the sinusoidal model may be systematically smaller than the actual $q$-values. Depending on how the modeling is done, inaccuracies in the estimation of $q$ may cause estimates of other system parameters to be systematically inaccurate as well. In any case, we confirm that the sinusoidal model should generally be sufficiently accurate to distinguish planetary companions from low mass stellar companions. 

\section{Observations of the HAT-P-7 System}
\label{sec:observations}
\indent The HAT-P-7 planetary system was discovered by the HATNet survey and was the second planet discovered in the Kepler field of view. The system is composed of an F-type star ($M_\star = 1.47 M_\odot$, $R_\star = 1.84 R_\odot$) and a gas giant planet ($M_p = 1.78 M_{Jup}$, $R_p = 1.36 R_{Jup}$) in a 2.2 day circular orbit \citep{2008ApJ...680.1450P}. In the Kepler bandpass, the star has a magnitude $K_p = 10.5$, relatively bright among Kepler targets. \citet{2009Sci...325..709B} analyzed the first ten days of Kepler data (quarter 0, Q0), detected a secondary eclipse depth of $130 \pm 11$ ppm, and estimated a dayside temperature of 2650 K. \citet{2010ApJ...710...97C} analyzed observations of HAT-P-7 b's secondary eclipse from the EPOXI mission and put upper limits on its depth at 0.055\%. They also analyzed Spitzer secondary eclipses taken throughout the IR and found brightness temperatures in the different bandpasses from 2250\, K to 3190 K.

\indent \citet{2010ApJ...713L.145W} discovered ellipsoidal variations in the Kepler Q1 data, with an amplitude of 37.3 ppm. They also estimated the planet's phase curve has an amplitude of 31.9 ppm and day- and nightside temperatures of 2885 and 2570 K, respectively. The discrepancy between the \citet{2009Sci...325..709B} result and the \citet{2010ApJ...713L.145W} result may arise from the consideration of ellipsoidal variation in the latter analysis and/or the inclusion of more data (Q1 data span 30 days, as compared to Q0's 10 days). 

\indent \citet{2009ApJ...703L..99W} conducted Rossiter-McLaughlin observations of HAT-P-7 and found the planet is in a near polar or even retrograde orbit about its star, with an angle between the stellar rotation and orbit normal vectors projected onto the sky-plane of $182.5 \pm 9.4^\circ$. They also pointed out that the unusually low projected rotational velocity of the star for its type and age suggests we are observing the star nearly pole-on. (Note that even for the estimated deprojected rotation velocity of $\sim$15 km/s from \citealt{2009ApJ...703L..99W}, HAT-P-7 is still a ``slow-rotator'' for the purposes of our linearized model.)

\begin{figure}
\includegraphics[width=0.45\textwidth]{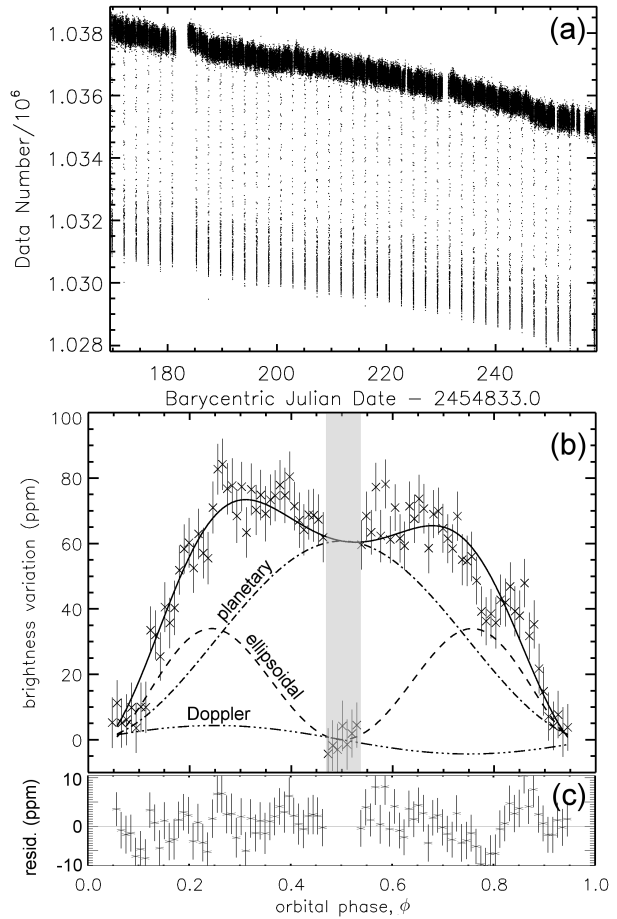}
\caption{(a) Kepler observations of HAT-P-7 system from Q2 with outliers filtered out (see text). (b) Observations from Q0-2 phased together and binned to 30-minute bins (Xs). Our best fit model curve (solid line) is also shown, the best-fit ellipsoidal variation is shown as a dashed curve, the planetary phase curve as a dash-dot line, and the Doppler flux variations (with $K_Z = 300$ m/s) as the dash-dot-dot-dot line. The planet's eclipse is highlighted in grey and is not fit by our model. Best fit parameters are shown in Table \ref{tbl:fit_params}. Our originally estimated uncertainties ($\sim$4 ppm) are re-scaled from the initial values by 1.8, the square root of the best-fit reduced $\chi^2 = 3.25$, to give uncertainties $\sim$8 ppm. (c) Residuals between the best-fit model and the data are nearly normally distributed about 0.}
\label{fig:fit_HP7}
\end{figure}

\indent For our analysis, we obtained the short-cadence (1-minute observing cadence) Pre-Search Data Conditioned (PDC) light curves from the MAST archive (\href{http://archive.stsci.edu/kepler/}{http://archive.stsci.edu/kepler/}), and we analyzed the Q0-2 data. (The other data publicly available at the time of our analysis from Q3 exhibited more complex systematic trends, so we did not include them in our analysis.) We first removed outlying photometric points that were explicitly flagged in the data files as anomalous by the Kepler team. We then binned the remaining data in 30-minute bins, calculated each bin's standard deviation, and threw out data points more than 4 standard deviations from the mean in each bin. The Q2 data filtered in these ways are shown in Figure \ref{fig:fit_HP7} (a). These data still clearly exhibit both long-term trends and correlated noise.

%
\indent We attempted to remove these trends. First, for the Q0 data, we masked out all the transits (5 transits) and fit a 4th-order polynomial to the remaining data. (3rd and 5th-order polynomials gave equivalent results within uncertainties.) We calculated the standard deviation $\sigma$ of residuals between data and the trend curve and dropped points that lay more than 4-$\sigma$ from the trend curve. We re-fit a 4th-order polynomial to these screened data and iterated this procedure until all 4-$\sigma$ outliers were removed. Then, we divided the data (including transits) by the final trend curve. We performed the same de-trending for the Q1 and Q2 data. Analyzing these three quarters together nearly quadruples the number of orbits examined over the analysis of \citet{2010ApJ...713L.145W} and significantly improves the accuracy of the estimated system parameters.

\indent After detrending the data, we phased and stacked them, assuming an orbital period of 2.204733 days \citep{2010ApJ...713L.145W}. We then binned the data into 30-minute wide bins, determined a median for each bin, and took 1.4826 $\times$ the median absolute deviation (MAD) as the standard deviation for each bin \citep{1969drea.book.....B}. We then threw out points in each bin more than 4-$\sigma$ from the median. We then took the median of the remaining data in each bin. For the uncertainties, we took 1.4826 $\times$ MAD divided by the square root of the number of points in each bin -- such uncertainties were typically 4 ppm. However, systematic trends or correlated noise still pervade the data \citep{2006MNRAS.373..231P}, producing scatter larger than 4 ppm. 

\indent To estimate the size of this scatter, we determined an initial best-fit model using a Levenberg-Marquadt algorithm \citep{2009ASPC..411..251M}, which gave a reduced $\chi^2 = 3.25$, indicating the scatter was indeed underestimated. We re-scaled the error bars by $\sqrt{3.25} = 1.8$, giving uncertainties $\sim$8 ppm.

\indent Finally, we calculated the overall normalization of the data by taking the mean of the data during the eclipse phase, when only the star is contributing flux. (Estimated variation of the system brightness during this phase is less than 0.1 ppm, and so variations in these data are dominated by intrinsic scatter.) We divided all the data through by this value (for Figure \ref{fig:fit_HP7} (b), we subtracted 1.0 from the data). These are the final data we analyzed and are shown in Figure \ref{fig:fit_HP7} (b) (with the transit near phase 0 and 1 masked out), along with our best model curve (see below). The contributions from ellipsoidal variations, the planetary phase curve, and Doppler flux variations are also shown.

\section{Analysis}
\label{sec:analysis}

\begin{table}
\centering
\caption{Fixed model parameters \label{tbl:fixed_params}}
\begin{tabular}{|c|c|}
\tableline
\textbf{param.} & \textbf{value}\\
\tableline
$R_\star/R_p$ & 12.85\tablenotemark{a}\\
\tableline
$a$ & 4.15\tablenotemark{a} \\
\tableline
$\omega_\star$ & 4.73 $\times 10^{-7} s^{-1}$\tablenotemark{b} \\
\tableline
$T_\star$ & 6350 K\tablenotemark{a}\\
\tableline
$[\frac{Fe}{H}]$ & 0.26\tablenotemark{b}\\
\tableline
log(g) & 4.07 (cm/s$^2$)\tablenotemark{b}\\
\tableline
$i$ & $83.1^\circ$\tablenotemark{a}\\
\tableline
P &2.204733 days\tablenotemark{a}\\
\tableline
$(\gamma_1, \gamma_2)$ & (0.314709, 0.312125)\tablenotemark{c}\\
\tableline
$\beta$ & 0.0705696\tablenotemark{c}\\
\tableline
\end{tabular}
\tablenotetext{1}{\citet{2010ApJ...713L.145W}}
\tablenotetext{2}{\citet{2008ApJ...680.1450P}}
\tablenotetext{3}{Determined from interpolation among the values in \citet{2011AA...529A..75C}}
\end{table}
\indent We conducted a suite of Markov-chain Monte Carlo (MCMC) analyses, using Gibbs sampling \citep{2005AJ....129.1706F} to fit the model parameters, $q$, $K_Z$, $F_0$, and $F_1$ (Table \ref{tbl:fit_params}). In some of the model fitting, though, we also held $K_Z$ constant, and the sum $F_0 + F_1$ was constrained by the eclipse depth (see below). We held all other parameters fixed for all modeling (Table \ref{tbl:fixed_params}). 

\indent The eclipse depth provides a constraint on the maximum of the planetary phase curve $D = F_0 + F_1$. We estimated the eclipse depth by fitting a straight line between the points on either side of the eclipse. Then, we took the eclipse depth to be the difference between that value and 1.0 (the normalized value during eclipse), giving $D = 61 \pm 3$ ppm. (The difference between the actual maximum in the planetary phase curve and the maximum estimated this way is considerably less than the scatter in the data.) We conducted two sequences of MCMC analyses: (1) with $D$ held constant (best-fit parameters for which are in the first column in Table \ref{tbl:fit_params}) and (2) allowing $D$ to float but with a $\chi^2$-penalty for departures from 61 ppm (second column). (For the latter analysis, $K_Z$ was allowed to float as well.) Comparison of the two sequences below highlights the degeneracy between constraints on the ellipsoidal variation and the planetary phase curve, but we focus our discussion on the analyses with $D =$ const since they give a $q$-value consistent with previous analyses.

\indent In theory, the ellipsoidal variation signal depends on the relative orientation of $\mathbf{\omega}_\star$ and the orbit normal vector. However, we tried different relative orientations and found the data cannot distinguish between an $\mathbf{\omega}_\star$ that points directly at the observer (parallel to $\mathbf{\hat{Z}}$ -- Figure \ref{fig:prob_geom}) and any other orientation allowed by other constraints \citep{2009ApJ...703L..99W}. Thus, we assumed the $\mathbf{\omega}_\star$ vector points directly at the observer in our modeling ($\mathbf{\omega_\star} || \mathbf{\hat{Z}}$). 

\indent For the MCMC fitting, we used 5 Markov chains, each with $5\times10^3$ links, which we show below suffices for good convergence of the model parameters. (We also conducted MCMC analyses with $5\times10^4$ links which confirmed the shorter chains had converged.) For each jump transition, we chose at random either 1 or 2 parameters to vary. We discarded the first 20\% of each chain. Otherwise, the resulting distributions of best-fit parameters might be skewed by our initial choice of parameter values. For sampling the parameter space, we took the Gaussian distribution suggested  by \citet{2005AJ....129.1706F} for the candidate transition probability, with a width $\beta$ for each parameter such that the fraction of accepted transitions was $\sim$0.25. (See Equation 12 in \citealt{2005AJ....129.1706F}.)

\indent We also checked that our analysis technique can accurately recover system parameters by generating several synthetic data sets designed to mimic the raw Kepler data with ellipsoidal variations,  planetary emission/reflection, Doppler flux variations, and the same gaps in time and scatter. We consistently recovered the assumed system parameters when they were recoverable (see discussion of $K_Z$ estimate below).

\begin{figure}
\includegraphics[width=0.5\textwidth]{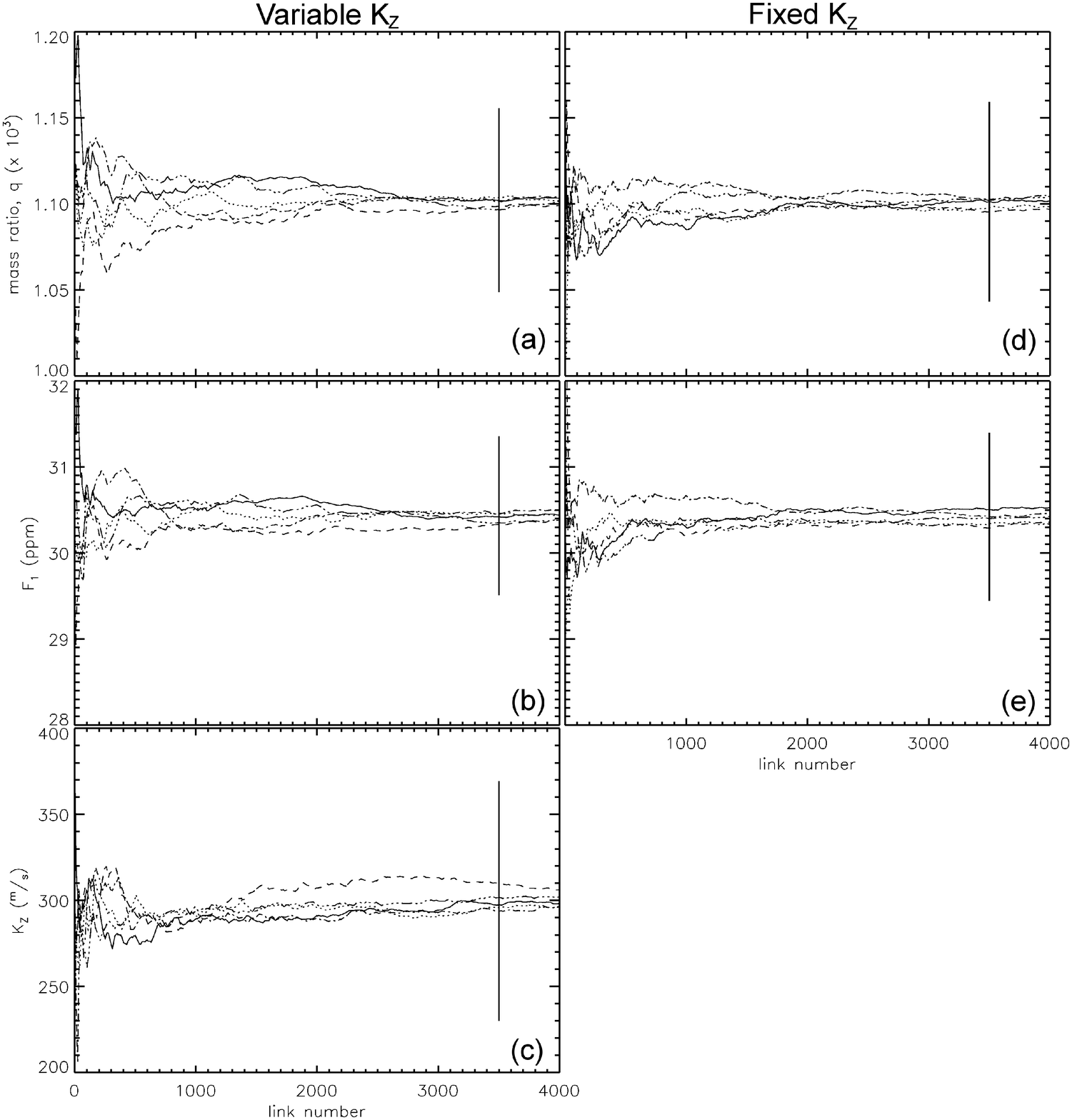}
\caption{The mean values for each parameter from each of the 5 Markov Chains as a function of link number (after the first 20\% of each chain was dropped). Each linestyle represents a different chain. The left column ((a)-(c)) shows the chains for which $K_Z$ is taken as a free parameter, while the right column ((d) \& (e)) shows the chains for which $K_Z = 213.5$ m/s. The vertical lines in each panel aligned with x = 3,500 represent the smallest standard deviation from among all the chains, and the means for all chains converge to within those deviations.}
\label{fig:check_conv}
\end{figure}

\indent For the sequence of analyses with the eclipse depth $D$ held constant, Figure \ref{fig:check_conv} illustrates the convergence of the mean for each of the 5 chains. For each parameter, the distribution from each chain provides a slightly different mean value, but the differences between the various mean values are all smaller than the smallest standard deviation for any one chain. For example, in Figure \ref{fig:check_conv} (a), the largest difference in the final mean $q$-value between different chains is $5.3\times10^{-6}$, while the smallest standard deviation from among all the chains is more than 10 times larger, indicating the chains have all converged within uncertainties. 

\begin{figure}
\includegraphics[width=0.45\textwidth]{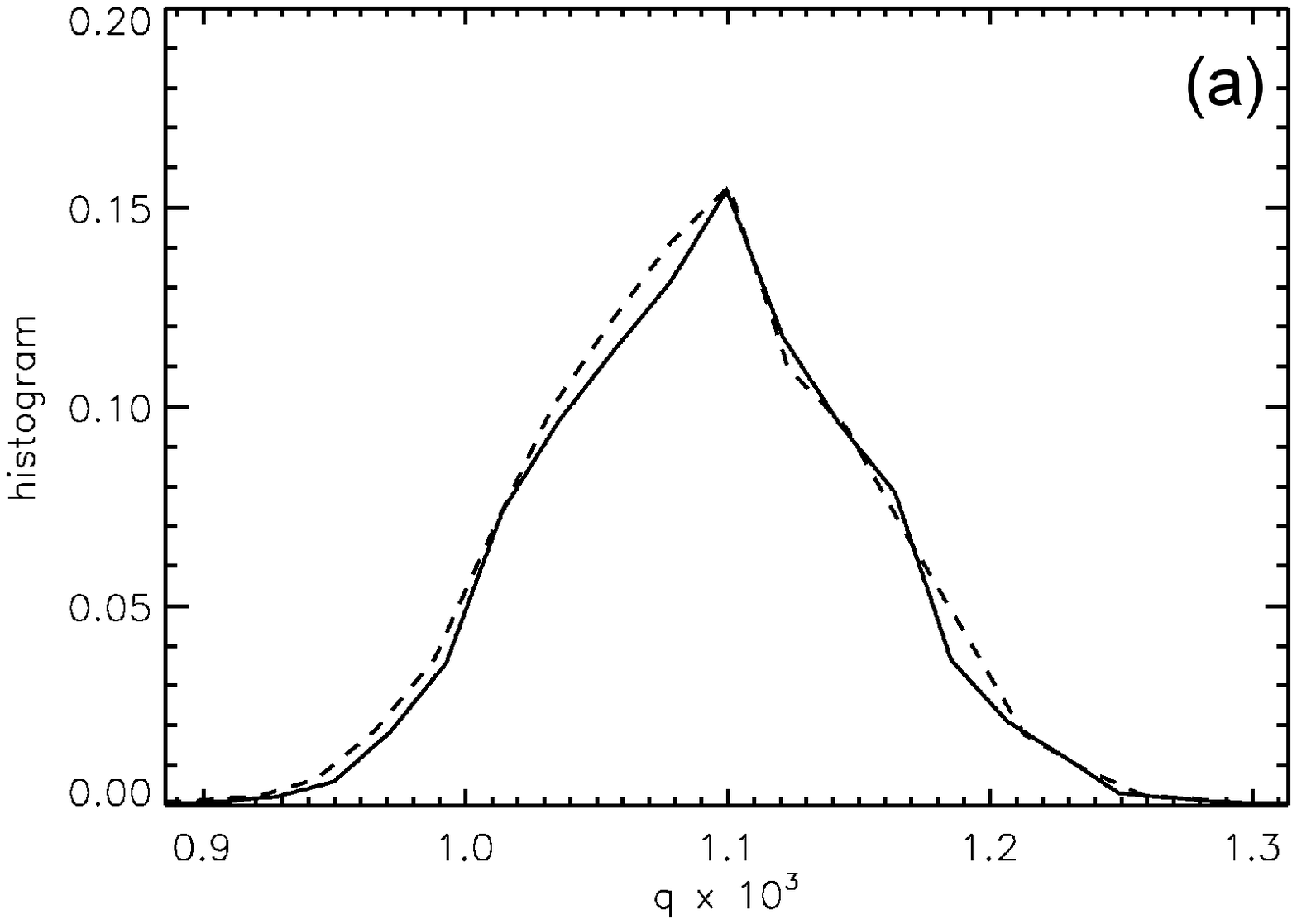}
\includegraphics[width=0.45\textwidth]{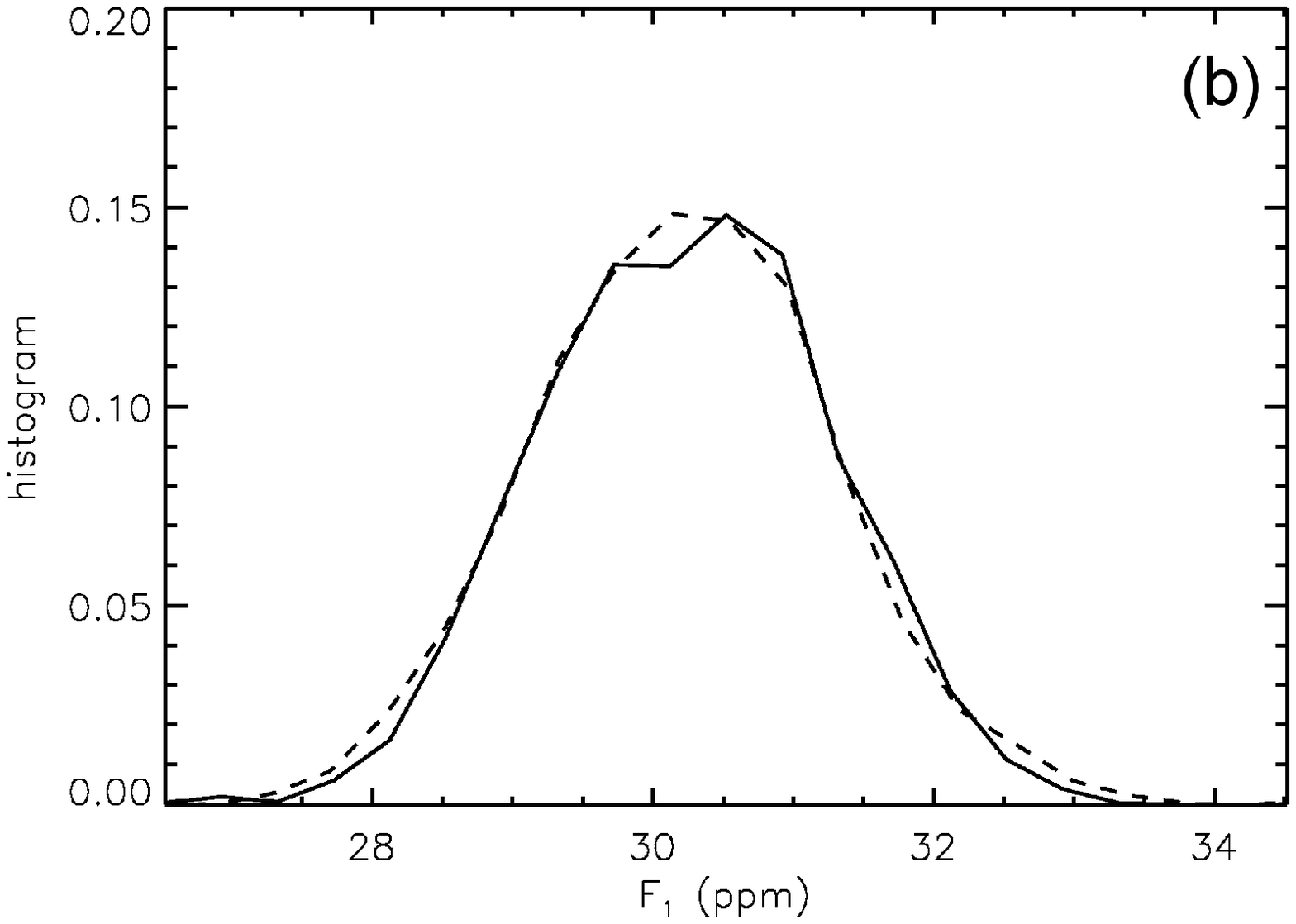}
\includegraphics[width=0.45\textwidth]{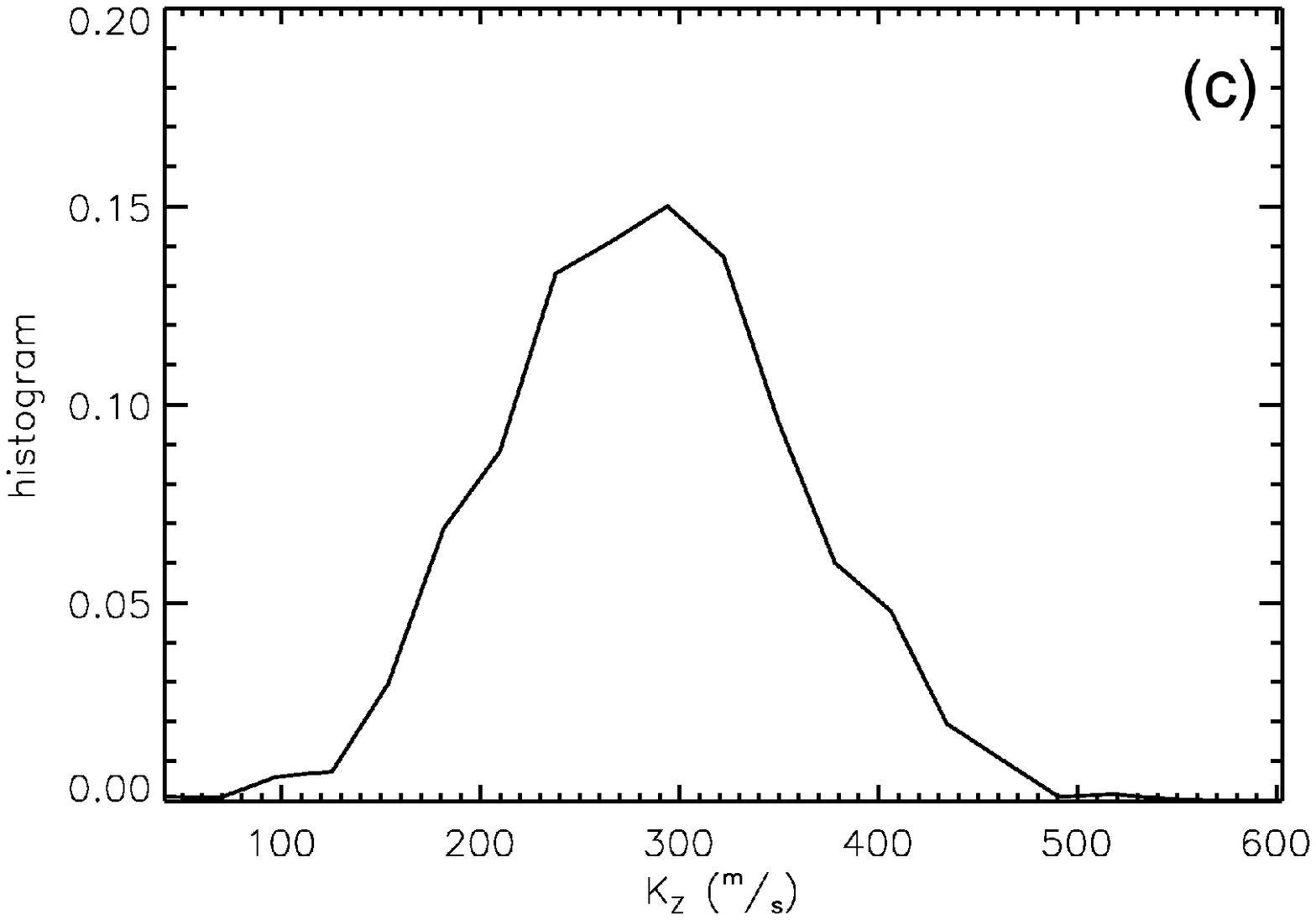}
\caption{The distributions of the model fit parameters resulting from our MCMC analysis, with $K_Z$ variable (solid) and fixed (dashed). (a) Mass ratio, $q$ -- For $M_\star = 1.47 M_{\odot}$ \citep{2008ApJ...680.1450P}, our best-fit $q$ gives $M_p = 1.62 M_{Jup}$. (b) Amplitude of the planetary phase curve, $F_1$ - The sum $F_0 + F_1$, the emission from the planet's dayside hemisphere, is held fixed at the estimated eclipse depth, 61 ppm. With the best-fit $F_1$ illustrated, $F_0 - F_1$, the emission from the planet's nightside hemisphere, is nearly 0. (c) Amplitude of the stellar reflex velocity, $K_Z$ -- Even for $K_Z = 300$ m/s, the signal from the Doppler flux variations has an amplitude of only about 4 ppm, twice as small as the intrinsic scatter in the data. Thus, the best-fit $K_Z$ is very sensitive to the scatter and has large uncertainties.}
\label{fig:params_dist}
\end{figure}

\indent Figure \ref{fig:params_dist} shows the distributions of best-fit parameters for the two sequences of model-fitting with constant $D$, one with $K_Z$ variable, the other with $K_Z$ fixed at 213.5 m/s \citep{2008ApJ...680.1450P}. The mean of each distribution is taken as the best-fit value, and the standard deviation is the uncertainty. Our best-fit $q$ ($(1.10\pm0.06)\times10^{-3}$) is smaller but consistent (within 2-$\sigma$) with that of \citet{2010ApJ...713L.145W} ($1.190\times10^{-3}$). Assuming $M_\star = 1.47 M_\odot$, our $q$-value corresponds to $M_p = 1.62 M_{Jup}$.

\indent \citet{2008ApJ...680.1450P} estimated $K_Z = 213.5$ m/s, but our best-fit value is $300 \pm 70$ m/s. This latter value corresponds to Doppler flux variations of only about 4 ppm, below the intrinsic scatter in the data. We conducted numerical tests to see whether we could, indeed, have recovered a $K_Z = 213.5$ m/s and found that we could only recover $K_Z$ for scatter less than or comparable to the Doppler signal. \citet{2010ApJ...713L.145W} estimated $K_Z = 212$ m/s by steering $K_Z$ toward 213.5 m/s via a $\chi^2$ penalty for deviations (although in Figure 3 of that study, the asymmetry that should result from the Doppler signal seems absent). In any case, as illustrated in Figure \ref{fig:params_dist} (a) and (b), the best-fit values for other parameters are insensitive to $K_Z$. 

\begin{figure}
\includegraphics[width=0.35\textwidth]{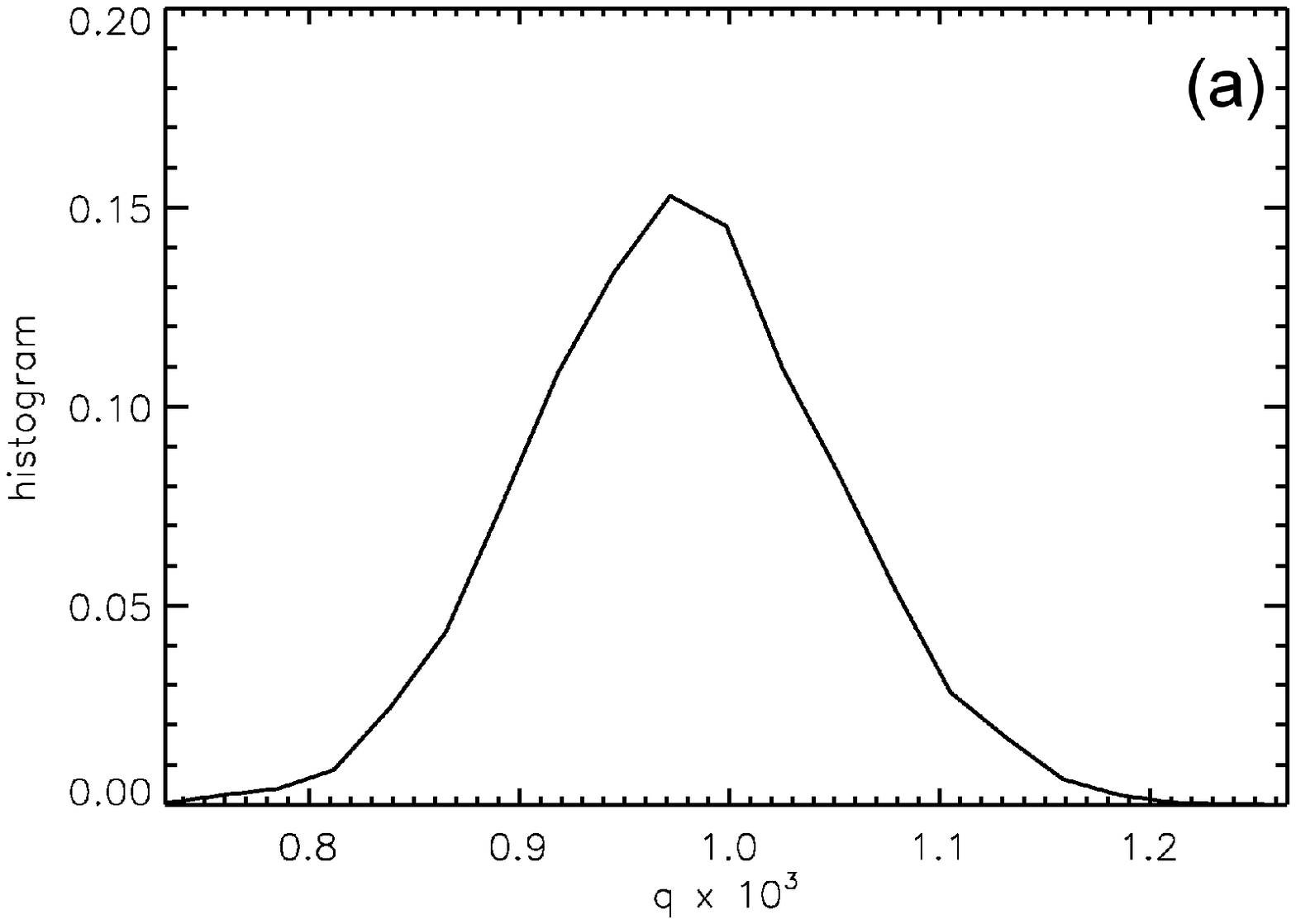}
\includegraphics[width=0.35\textwidth]{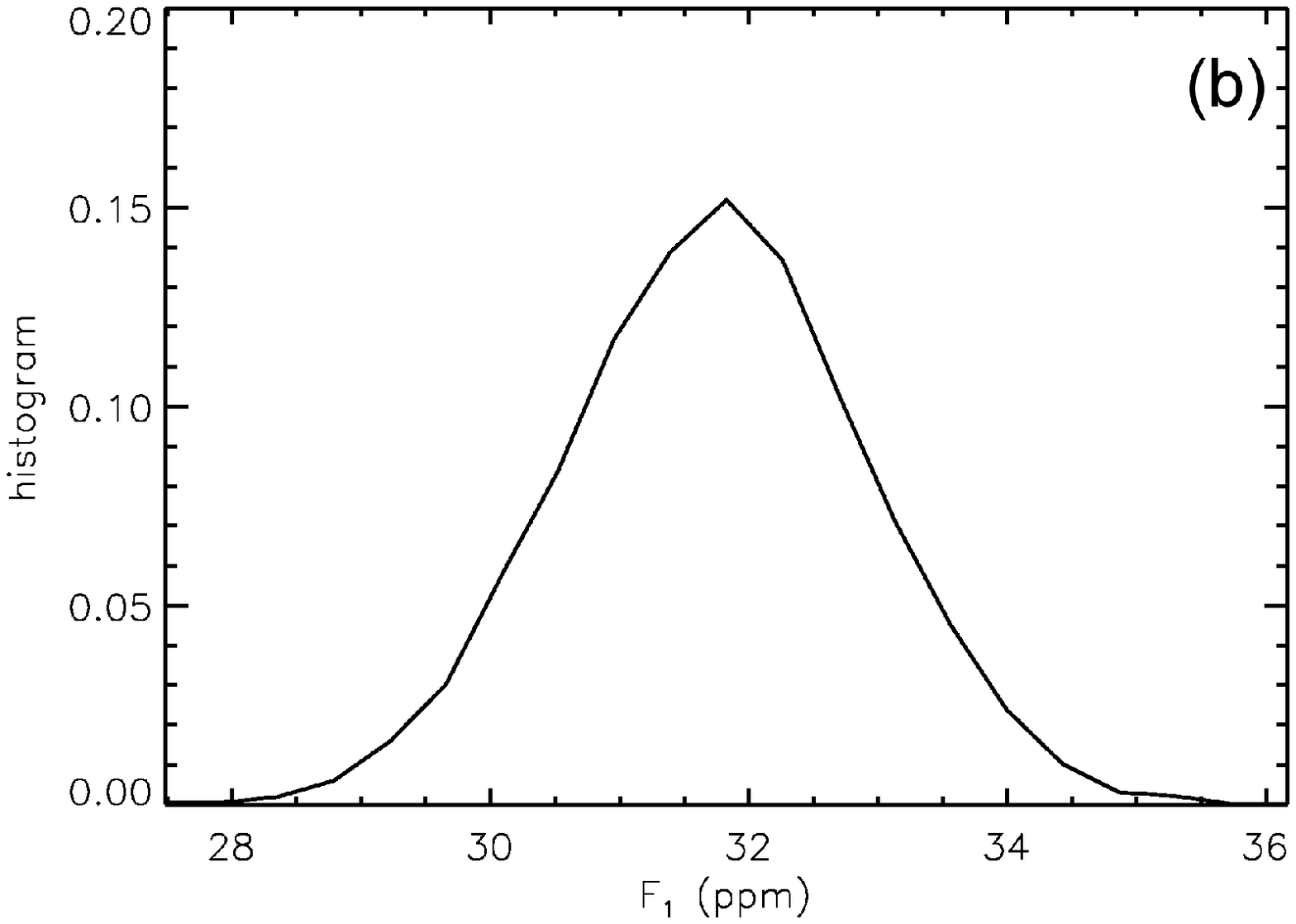}
\includegraphics[width=0.35\textwidth]{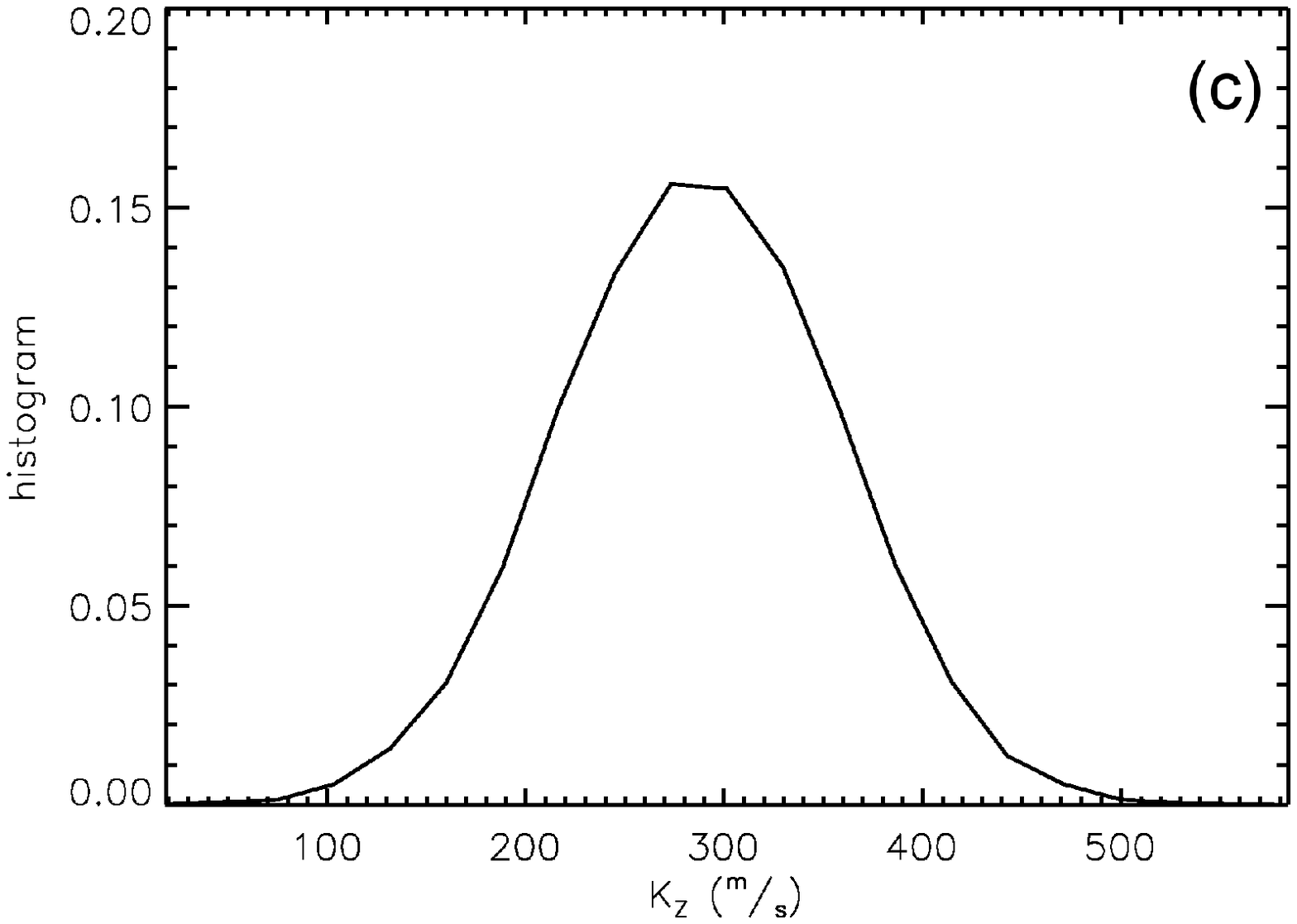}
\includegraphics[width=0.35\textwidth]{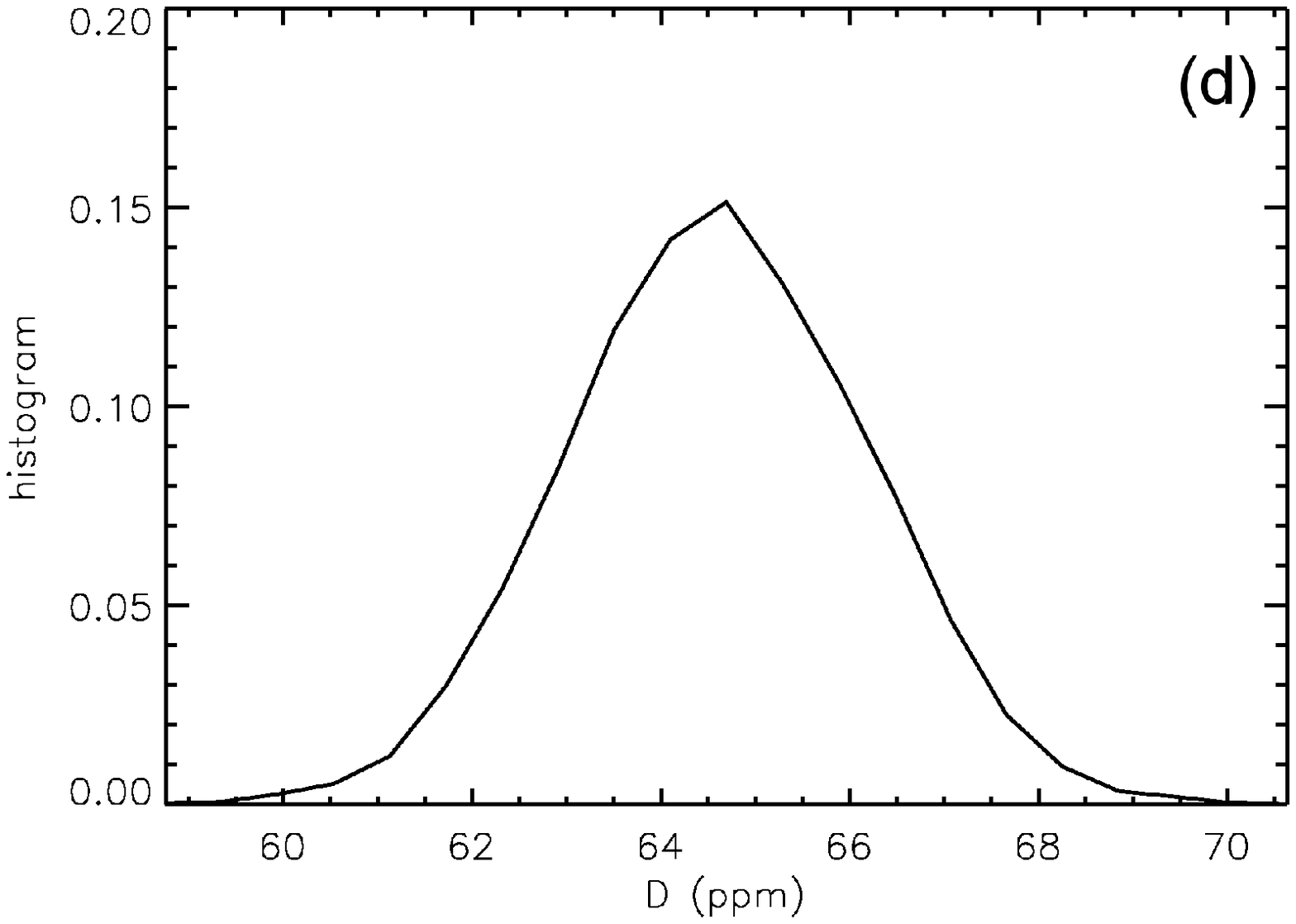}
\caption{The distributions of the model fit parameters resulting from our MCMC analysis, with $K_Z$ variable. The eclipse depth $D$ is also allowed to float but with a $\chi^2$-penalty for departures from 61 ppm. (a) Mass ratio, $q$ -- The MCMC analysis drives $q$ to smaller values when $D$ is allowed to float, producing a best-fit $q = 0.99 \pm 0.07 \times 10^{-3}$. (b) Amplitude of the planetary phase curve, $F_1$ - This value also goes up as $D$ increases, producing a best-fit $F_1 = 32 \pm 1$ ppm. (c) Amplitude of the stellar reflex velocity, $K_Z$ -- This parameter is essentially unchanged and has a best-fit value $K_Z = 300 \pm 70$ m/s. (d) Eclipse depth $D$ -- The MCMC routine drives this parameter to larger values than our best estimate, producing a best-fit value $D = 65 \pm 2$ ppm.}
\label{fig:params_dist_floatD}
\end{figure}

\indent As discussed above, our estimated eclipse depth for the planet is $61 \pm 3$ ppm, as compared to the eclipse depth of 85.8 ppm from \citet{2010ApJ...713L.145W}. This discrepancy arises from our use of more data than used in that previous study. A preliminary analysis of Q1 data alone yielded an eclipse depth similar to that of \citet{2010ApJ...713L.145W}. Assuming the planet's dayside emits as a uniform blackbody, our depth corresponds to a dayside temperature of $2680^{+10}_{-20}$ K, which is almost 200 K smaller than the average dayside temperature from \citet{2010ApJ...713L.145W}. The disagreement with the eclipse depth from \citet{2009Sci...325..709B} probably arises for similar reasons. As a further confirmation of our estimate, \citet{2012A&A...538A...4M} conducted an analysis of some of the same data as we and found a similar eclipse depth (see their Figure 7).

\indent For our analyses with fixed $D$, we found that the nightside emission $F_0 - F_1 \simeq 0$, as compared to the 22.1 ppm estimated by \citet{2010ApJ...713L.145W}. Partly, this disagreement is due to the fact that we do not explicitly analyze the transit phase, while \citet{2010ApJ...713L.145W} do, and partly, it is due to our choice of planetary phase function: \citet{2010ApJ...713L.145W} chose a planetary emission/reflection relationship that produces a shallower drop off in planetary flux than our function as $\phi$ departs from 0.5. Both estimates for the nightside emission are model-dependent, though.

\indent Figure \ref{fig:params_dist_floatD} illustrates the results of the MCMC analysis in which $D$ was allowed to float. (Note: convergence of the model parameters for this analysis required $5\times10^4$ links in each chain.) Our MCMC analysis drives $q$ to smaller values and $D$ to larger values than when $D$ is held fixed. Unfortunately, because we re-scaled our uncertainties to force $\chi^2 \sim 1$, we cannot use the $\chi^2$-values from the different MCMC sequences to determine whether letting $D$ float provides a statistically better model fit. However, the fact that the $q$-value for the sequence with fixed $D$ more closely matches previous constraints suggests that is the more appropriate model. In any case, Table \ref{tbl:fit_params} shows the planetary parameters corresponding to the best-fit values for variable $D$.

\indent We also conducted numerical experiments for which we created synthetic datasets with the same best-fit parameters produced by the previous MCMC analysis (column 1 in Table \ref{tbl:fit_params}). We added Gaussian noise to these synthetic datasets, with a scatter of 8 ppm. We applied the same MCMC analysis in which we allowed $D$ to float, and the analysis would often drive $q$ to smaller values and $D$ to larger values than assumed, depending on exactly where the noisy data points ended up. These results highlight the degeneracy between the best-fit $q$ and planetary phase function and show that it can depend sensitively on the scatter in the data.

\indent The dependence of the derived $q$-value on the assumed planetary phase function has been considered by \citet{2012A&A...538A...4M}. For any planetary phase function that is symmetric about $\phi = 0.5$, there will necessarily be some degeneracy between the solution for the phase function and $q$. For example, a model fit to light emerging from a planet-star system exhibiting ellipsoidal variations can enhance the peaks near $\phi = 0.25$ and $0.75$ by increasing the baseline planetary flux ($F_0$), subject to constraints on the eclipse depth, or by increasing $q$. Additional constraints on the planetary emission from other phases cannot completely remove this degeneracy. Additional degeneracies between, for example, the planetary emission and the transit parameters should emerge from analysis of the transit phase. 

\begin{figure}
\includegraphics[width=0.45\textwidth]{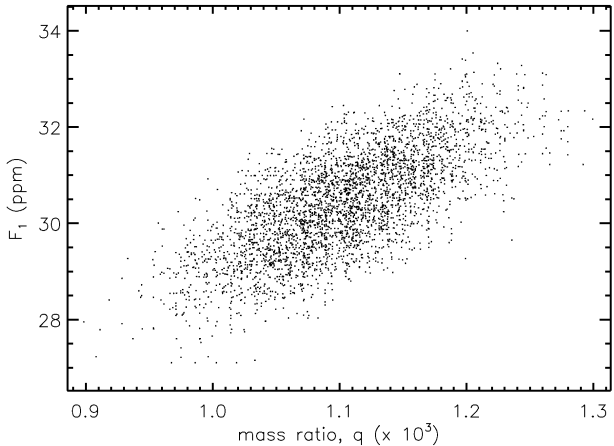}
\caption{The values for mass ratio $q$ and the amplitude of the planetary phase function $F_1$ sampled during the MCMC analysis. The strong positive correlation between these parameters arises because an increase in flux from a planet-star system when the planet is near quadrature can be attributed to increasing $F_0$/reducing $F_1$ (their sum is constrained) and reducing $q$ or vice versa.}
\label{fig:q_F1_degen}
\end{figure}

\indent Even for $D = $ const., this degeneracy persists. Figure \ref{fig:q_F1_degen} illustrates the degeneracy between $q$ and $F_1$ and shows the values sampled by the MCMC routine (with variable $K_Z$) when $D$ is held constant. Since $D = F_0 + F_1 = $ const., an increase in the signal from the system when the planet is near quadrature can be attributed to either increased $q$ or increased $F_0$/decreased $F_1$. A similar degeneracy does not exist for $K_Z$ since the Doppler flux variations aren't symmetric about $\phi = 0.5$. Taken altogether, our results show that, while Kepler and CoRoT data may provide important constraints on planetary albedo and energy budget, constraining these properties requires including ellipsoidal variations. The different contributions cannot be completely disentangled. Future work should consider more completely the influence of alternative planetary phase functions.

\section{Discussion and Conclusions}
\label{sec:disc}

\indent We have developed a new model for ellipsoidal variations induced by close-in planets, the EVIL-MC model. EVIL-MC employs several approximations suited for planet-star systems and is thus computationally more efficient than other more general models and more accurate than simpler, semi-analytic models. For example, the W-D model takes about 0.5 s to run each of the examples in Section \ref{sec:comp_models}, while our EVIL-MC model (in IDL) runs in less than 0.05 s. Also, after our HAT-P-7 data were detrended and binned, we performed the entire suite of MCMC calculations (25,000 evaluations of the EVIL-MC model) described in Section \ref{sec:analysis} in about 20 minutes. This increase in efficiency makes our model well-suited for analyzing the mountain of Kepler and CoRoT data still pouring in. 

\indent The EVIL-MC model has some important limitations. It is not designed for systems with a mass ratio $q \sim 1$ since tidal distortions are not small for those systems, although our  comparison to the more general W-D model shows agreement at about 2\% even for large $q$-values (Section \ref{sec:comp_models}). 

\indent The EVIL-MC model may not be sufficiently accurate for rapidly rotating stars where the rotational oblateness is large. Rotationally induced gravity darkening at the equators of such stars may imprint a discernible signature on the transit light curves of companion planets, analysis of which can reveal the misalignment between a planet's orbit and the stellar equator. Such analyses have been conducted for the KOI-13.01 system \citep{2011ApJS..197...10B, 2011ApJ...736L...4S}.  

\indent EVIL-MC also does not currently include the transit and eclipse phases for a planetary system, and a future version will also include these phases. However, a preliminary analysis shows that tidal distortion has a negligible ($<$ 0.1 ppm) influence on the transit light curves for typical planetary systems.

\indent Accurate determination of planetary phase curves from Kepler and CoRoT data requires consideration of the ellipsoidal variations, and there can be degeneracies between the contributions from the ellipsoidal variation and the planetary phase curve. Planetary phase curves are diagnostic of atmospheric temperatures and dynamics, and a complex story of coupled chemistry, dynamics, and radiation is emerging, motivated largely by IR observations of planetary phase curves and eclipse depths (see, e.g., \citealp{2010ApJ...720.1569K}). Results from the Kepler and CoRoT missions will add to this picture and, when combined with Spitzer observations, will give a much fuller picture of the atmospheric energy budgets of close-in planets.

\indent From our analysis, we can draw some tentative conclusions regarding HAT-P-7 b's atmosphere. Given its proximity to its host star, the planet is probably tidally locked, and the same side of the planet always faces the star \citep{2008ApJ...678.1396J}. Consequently, atmospheric circulation is required to transport stellar heating from the day- to the nightside. Our estimated dayside emission 61 ppm corresponds to a brightness temperature of 2680 K. Our estimated minimum for the planet's nightside emission $F_0 - F_1$ lies below the sensitivity of our analysis, $\sim 4$ ppm, suggesting the nightside brightness temperature in the Kepler band is less than 1970 K.

\indent This result might indicate much of the stellar heating on the dayside is radiated to space before it can be transported to the nightside. This result is also qualitatively consistent with models of the hottest close-in planets \citep{2008ApJ...678.1419F} and with analyses that suggest HAT-P-7 b has an atmospheric thermal inversion \citep{2010ApJ...710...97C}, which is often correlated with a high atmospheric temperature for close-in planets \citep{2010ApJ...720.1569K}. However, determining the precise implications of this result for the atmospheric circulation requires detailed modeling. It is worth noting that the day-night brightness temperature contrast inferred here (~710 K) is similar to that inferred for WASP-12 b \citep{2012ApJ...747...82C} but greater than those of cooler hot Jupiters, including HD 189733 b, \citep{2007Natur.447..183K}, HD 209458b \citep{2007MNRAS.379..641C}, and HD 149026b \citep{2009ApJ...703..769K} (although the latter has an error bar that does allow relatively large values). These measurements are not all at the same wavelength, which complicates the interpretation. An additional complication is that our eclipse depth may also include contributions from atmospheric scattering of light by clouds, although estimated optical albedos of hot Jupiters are highly uncertain \citep{2008ApJ...689.1345R, 2011ApJ...729...54C}. 

\indent To help place our results regarding HAT-P-7 b in context, we ran some preliminary dynamical+radiative calculations using the SPARC model \citep{2009ApJ...699..564S}. The HAT-P-7b model atmospheres were constructed assuming a solar metallicity atmosphere in thermochemical equilibrium for cases with and without TiO, which can absorb stellar radiation high in the atmosphere and produce a temperature inversion \citep{2010ApJ...710...97C}. 

\indent For the model with TiO in the atmosphere, stellar heating is deposited higher in the atmosphere, where the timescale for radiation of stellar heating to space is relatively short, and consequently the model predicts a large day-night contrast: a dayside emission of 74 ppm and a nightside emission of only 1 ppm. For the model without TiO, stellar heating is deposited deeper in the atmosphere, where the radiative timescale is relatively long, and so the model predicts a smaller day-night contrast: a dayside emission of 52 ppm and nightside emission of 9 ppm. Comparison with our observations suggests HAT-P-7 b's real atmosphere might occupy a point in parameter space somewhere between these models. Our results here suggest there are still important, unanswered questions about HAT-P-7 b. The planet is one of the hottest hot Jupiters known and orbits one of the brightest Kepler targets, and so further study of the planet may prove particularly useful for understanding hot Jupiter atmospheres.

\indent Kepler and CoRoT observations will provide numerous opportunities for similar phase curve analyses. The closer a planet to its host star, the more stellar radiation it will receive, probably leading to greater reflection and/or thermal emission. The tidal distortion and ellipsoidal variation of its host star could also be larger. Accurate determination of phase curves for the closest-in planets will therefore require inclusion of the stellar ellipsoidal variation. Phase curve fitting without it may produce erroneous results. By contrast, ellipsoidal variation of a star has less influence on determination of planetary phase curves from Spitzer observations because the planet-star contrast for most extrasolar systems is much larger in the IR. 

\indent Ellipsoidal variation analysis may provide other key information about extrasolar systems. For example, \citet{2003ApJ...588L.117L} first suggested Doppler flux variations would be an important source of variability for Kepler observations. Equation \ref{eqn:reflex_vel} shows that Doppler variations (or radial velocity observations) has a different dependence on the system parameters than ellipsoidal variations. Potentially, transit observations would give the orbital period $P$ and orbital inclination $\sin i$, ellipsoidal variations would give the planet-star mass ratio $q$, leaving only the stellar mass unknown in Equation \ref{eqn:reflex_vel}.  

\indent Single planets close enough to their star to induce measurable ellipsoidal variations are likely to have negligible orbital eccentricities. However, if primordial eccentricities remain or interactions with other planets keep eccentricities non-zero (and the planet's semi-major axis isn't aligned along the line of sight), the planet-star orbital separation will be different at each quadrature. Consequently, the ellipsoidal variations at one quadrature may exceed that at the opposite quadrature, and the difference may help constrain the orbital orientation and eccentricity \citep{2012A&A...538A...4M}.

\indent Moreover, given the number of planets likely to be discovered by the Kepler and CoRoT missions, follow-up resources to determine the system parameters will be limited, so the ability to determine some of the parameters from mission photometry alone will be a tremendous boon. Thus, ellipsoidal variation analysis of Kepler and CoRoT systems promises to reveal a unique wealth of information.

\acknowledgments
The authors gratefully acknowledge useful conversations with Phil Arras, Jan Budaj, Nick Cowan, Maki Hattori, Dimitris Mislis, Jerome Orosz, Darin Ragozzine, William Welsh, and Robert E. Wilson. Input from an anonymous referee also greatly improved the paper.

%
%
\bibliography{myrefs}

\begin{thebibliography}{52}
\expandafter\ifx\csname natexlab\endcsname\relax\def\natexlab#1{#1}\fi

\bibitem[{{Arras} {et~al.}(2012){Arras}, {Burkart}, {Quataert}, \&
  {Weinberg}}]{2012MNRAS.tmp.2682A}
{Arras}, P., {Burkart}, J., {Quataert}, E., \& {Weinberg}, N.~N. 2012, \mnras,
  2682

\bibitem[{{Barnes} {et~al.}(2011){Barnes}, {Linscott}, \&
  {Shporer}}]{2011ApJS..197...10B}
{Barnes}, J.~W., {Linscott}, E., \& {Shporer}, A. 2011, \apjs, 197, 10

\bibitem[{{Bevington}(1969)}]{1969drea.book.....B}
{Bevington}, P.~R. 1969, {Data reduction and error analysis for the physical
  sciences}, ed. {Bevington, P.~R.}

\bibitem[{{Borucki} {et~al.}(2009){Borucki}, {Koch}, {Jenkins}, {Sasselov},
  {Gilliland}, {Batalha}, {Latham}, {Caldwell}, {Basri}, {Brown},
  {Christensen-Dalsgaard}, {Cochran}, {DeVore}, {Dunham}, {Dupree}, {Gautier},
  {Geary}, {Gould}, {Howell}, {Kjeldsen}, {Lissauer}, {Marcy}, {Meibom},
  {Morrison}, \& {Tarter}}]{2009Sci...325..709B}
{Borucki}, W.~J., {et~al.} 2009, Science, 325, 709

\bibitem[{{Borucki} {et~al.}(2011){Borucki}, {Koch}, {Basri}, {Batalha},
  {Brown}, {Bryson}, {Caldwell}, {Christensen-Dalsgaard}, {Cochran}, {DeVore},
  {Dunham}, {Gautier}, {Geary}, {Gilliland}, {Gould}, {Howell}, {Jenkins},
  {Latham}, {Lissauer}, {Marcy}, {Rowe}, {Sasselov}, {Boss}, {Charbonneau},
  {Ciardi}, {Doyle}, {Dupree}, {Ford}, {Fortney}, {Holman}, {Seager},
  {Steffen}, {Tarter}, {Welsh}, {Allen}, {Buchhave}, {Christiansen}, {Clarke},
  {Das}, {D{\'e}sert}, {Endl}, {Fabrycky}, {Fressin}, {Haas}, {Horch},
  {Howard}, {Isaacson}, {Kjeldsen}, {Kolodziejczak}, {Kulesa}, {Li}, {Lucas},
  {Machalek}, {McCarthy}, {MacQueen}, {Meibom}, {Miquel}, {Prsa}, {Quinn},
  {Quintana}, {Ragozzine}, {Sherry}, {Shporer}, {Tenenbaum}, {Torres},
  {Twicken}, {Van Cleve}, {Walkowicz}, {Witteborn}, \&
  {Still}}]{2011ApJ...736...19B}
---. 2011, \apj, 736, 19

\bibitem[{{Budaj}(2011)}]{2011AJ....141...59B}
{Budaj}, J. 2011, \aj, 141, 59

\bibitem[{{Christiansen} {et~al.}(2010){Christiansen}, {Ballard},
  {Charbonneau}, {Madhusudhan}, {Seager}, {Holman}, {Wellnitz}, {Deming},
  {A'Hearn}, \& {the EPOXI Team}}]{2010ApJ...710...97C}
{Christiansen}, J.~L., {et~al.} 2010, \apj, 710, 97

\bibitem[{{Claret}(2000)}]{2000AA...359..289C}
{Claret}, A. 2000, \aap, 359, 289

\bibitem[{{Claret} \& {Bloemen}(2011)}]{2011AA...529A..75C}
{Claret}, A., \& {Bloemen}, S. 2011, \aap, 529, A75

\bibitem[{{Cowan} \& {Agol}(2008)}]{2008ApJ...678L.129C}
{Cowan}, N.~B., \& {Agol}, E. 2008, \apjl, 678, L129

\bibitem[{{Cowan} \& {Agol}(2011)}]{2011ApJ...729...54C}
---. 2011, \apj, 729, 54

\bibitem[{{Cowan} {et~al.}(2007){Cowan}, {Agol}, \&
  {Charbonneau}}]{2007MNRAS.379..641C}
{Cowan}, N.~B., {Agol}, E., \& {Charbonneau}, D. 2007, \mnras, 379, 641

\bibitem[{{Cowan} {et~al.}(2012){Cowan}, {Machalek}, {Croll}, {Shekhtman},
  {Burrows}, {Deming}, {Greene}, \& {Hora}}]{2012ApJ...747...82C}
{Cowan}, N.~B., {Machalek}, P., {Croll}, B., {Shekhtman}, L.~M., {Burrows}, A.,
  {Deming}, D., {Greene}, T., \& {Hora}, J.~L. 2012, \apj, 747, 82

\bibitem[{{Deming} {et~al.}(2011){Deming}, {Sada}, {Jackson}, {Peterson},
  {Agol}, {Knutson}, {Jennings}, {Haase}, \& {Bays}}]{2011ApJ...740...33D}
{Deming}, D., {et~al.} 2011, \apj, 740, 33

\bibitem[{{Drake}(2003)}]{2003ApJ...589.1020D}
{Drake}, A.~J. 2003, \apj, 589, 1020

\bibitem[{{Ford}(2005)}]{2005AJ....129.1706F}
{Ford}, E.~B. 2005, \aj, 129, 1706

\bibitem[{{Fortney} {et~al.}(2008){Fortney}, {Lodders}, {Marley}, \&
  {Freedman}}]{2008ApJ...678.1419F}
{Fortney}, J.~J., {Lodders}, K., {Marley}, M.~S., \& {Freedman}, R.~S. 2008,
  \apj, 678, 1419

\bibitem[{{Jackson} {et~al.}(2009){Jackson}, {Barnes}, \&
  {Greenberg}}]{2009ApJ...698.1357J}
{Jackson}, B., {Barnes}, R., \& {Greenberg}, R. 2009, \apj, 698, 1357

\bibitem[{{Jackson} {et~al.}(2008){Jackson}, {Greenberg}, \&
  {Barnes}}]{2008ApJ...678.1396J}
{Jackson}, B., {Greenberg}, R., \& {Barnes}, R. 2008, \apj, 678, 1396

\bibitem[{{Knutson} {et~al.}(2009){Knutson}, {Charbonneau}, {Cowan}, {Fortney},
  {Showman}, {Agol}, \& {Henry}}]{2009ApJ...703..769K}
{Knutson}, H.~A., {Charbonneau}, D., {Cowan}, N.~B., {Fortney}, J.~J.,
  {Showman}, A.~P., {Agol}, E., \& {Henry}, G.~W. 2009, \apj, 703, 769

\bibitem[{{Knutson} {et~al.}(2010){Knutson}, {Howard}, \&
  {Isaacson}}]{2010ApJ...720.1569K}
{Knutson}, H.~A., {Howard}, A.~W., \& {Isaacson}, H. 2010, \apj, 720, 1569

\bibitem[{{Knutson} {et~al.}(2007){Knutson}, {Charbonneau}, {Allen}, {Fortney},
  {Agol}, {Cowan}, {Showman}, {Cooper}, \& {Megeath}}]{2007Natur.447..183K}
{Knutson}, H.~A., {et~al.} 2007, \nat, 447, 183

\bibitem[{{Kopal}(1959)}]{1959cbs..book.....K}
{Kopal}, Z. 1959, {Close binary systems}

\bibitem[{{Levrard} {et~al.}(2009){Levrard}, {Winisdoerffer}, \&
  {Chabrier}}]{2009ApJ...692L...9L}
{Levrard}, B., {Winisdoerffer}, C., \& {Chabrier}, G. 2009, \apjl, 692, L9

\bibitem[{{Loeb} \& {Gaudi}(2003)}]{2003ApJ...588L.117L}
{Loeb}, A., \& {Gaudi}, B.~S. 2003, \apjl, 588, L117

\bibitem[{{Mandel} \& {Agol}(2002)}]{2002ApJ...580L.171M}
{Mandel}, K., \& {Agol}, E. 2002, \apjl, 580, L171

\bibitem[{{Markwardt}(2009)}]{2009ASPC..411..251M}
{Markwardt}, C.~B. 2009, in Astronomical Society of the Pacific Conference
  Series, Vol. 411, Astronomical Data Analysis Software and Systems XVIII, ed.
  {D.~A.~Bohlender, D.~Durand, \& P.~Dowler}, 251

\bibitem[{{Mazeh} \& {Faigler}(2010)}]{2010AA...521L..59M}
{Mazeh}, T., \& {Faigler}, S. 2010, \aap, 521, L59

\bibitem[{{Mazeh} {et~al.}(2011){Mazeh}, {Nachmani}, {Sokol}, {Faigler}, \&
  {Zucker}}]{2011arXiv1110.3512M}
{Mazeh}, T., {Nachmani}, G., {Sokol}, G., {Faigler}, S., \& {Zucker}, S. 2011,
  ArXiv e-prints

\bibitem[{{Mislis} {et~al.}(2012){Mislis}, {Heller}, {Schmitt}, \&
  {Hodgkin}}]{2012A&A...538A...4M}
{Mislis}, D., {Heller}, R., {Schmitt}, J.~H.~M.~M., \& {Hodgkin}, S. 2012,
  \aap, 538, A4

\bibitem[{{Morris}(1985)}]{1985ApJ...295..143M}
{Morris}, S.~L. 1985, \apj, 295, 143

\bibitem[{{Murray} \& {Dermott}(1999)}]{1999ssd..book.....M}
{Murray}, C.~D., \& {Dermott}, S.~F. 1999, {Solar system dynamics}

\bibitem[{{Orosz} \& {Hauschildt}(2000)}]{2000AA...364..265O}
{Orosz}, J.~A., \& {Hauschildt}, P.~H. 2000, \aap, 364, 265

\bibitem[{{P{\'a}l} {et~al.}(2008){P{\'a}l}, {Bakos}, {Torres}, {Noyes},
  {Latham}, {Kov{\'a}cs}, {Marcy}, {Fischer}, {Butler}, {Sasselov}, {Sip{\H
  o}cz}, {Esquerdo}, {Kov{\'a}cs}, {Stefanik}, {L{\'a}z{\'a}r}, {Papp}, \&
  {S{\'a}ri}}]{2008ApJ...680.1450P}
{P{\'a}l}, A., {et~al.} 2008, \apj, 680, 1450

\bibitem[{{Pfahl} {et~al.}(2008){Pfahl}, {Arras}, \&
  {Paxton}}]{2008ApJ...679..783P}
{Pfahl}, E., {Arras}, P., \& {Paxton}, B. 2008, \apj, 679, 783

\bibitem[{{Pont} {et~al.}(2006){Pont}, {Zucker}, \&
  {Queloz}}]{2006MNRAS.373..231P}
{Pont}, F., {Zucker}, S., \& {Queloz}, D. 2006, \mnras, 373, 231

\bibitem[{{Rowe} {et~al.}(2008){Rowe}, {Matthews}, {Seager}, {Miller-Ricci},
  {Sasselov}, {Kuschnig}, {Guenther}, {Moffat}, {Rucinski}, {Walker}, \&
  {Weiss}}]{2008ApJ...689.1345R}
{Rowe}, J.~F., {et~al.} 2008, \apj, 689, 1345

\bibitem[{{Rybicki} \& {Lightman}(1979)}]{1979rpa..book.....R}
{Rybicki}, G.~B., \& {Lightman}, A.~P. 1979, {Radiative processes in
  astrophysics}

\bibitem[{{Showman} {et~al.}(2009){Showman}, {Fortney}, {Lian}, {Marley},
  {Freedman}, {Knutson}, \& {Charbonneau}}]{2009ApJ...699..564S}
{Showman}, A.~P., {Fortney}, J.~J., {Lian}, Y., {Marley}, M.~S., {Freedman},
  R.~S., {Knutson}, H.~A., \& {Charbonneau}, D. 2009, \apj, 699, 564

\bibitem[{{Shporer} {et~al.}(2011){Shporer}, {Jenkins}, {Rowe}, {Sanderfer},
  {Seader}, {Smith}, {Still}, {Thompson}, {Twicken}, \&
  {Welsh}}]{2011AJ....142..195S}
{Shporer}, A., {et~al.} 2011, \aj, 142, 195

\bibitem[{{Southworth} {et~al.}(2004){Southworth}, {Maxted}, \&
  {Smalley}}]{2004MNRAS.351.1277S}
{Southworth}, J., {Maxted}, P.~F.~L., \& {Smalley}, B. 2004, \mnras, 351, 1277

\bibitem[{{Szab{\'o}} {et~al.}(2011){Szab{\'o}}, {Szab{\'o}}, {Benk{\H o}},
  {Lehmann}, {Mez{\H o}}, {Simon}, {K{\H o}v{\'a}ri}, {Hodos{\'a}n},
  {Reg{\'a}ly}, \& {Kiss}}]{2011ApJ...736L...4S}
{Szab{\'o}}, G.~M., {et~al.} 2011, \apjl, 736, L4

\bibitem[{{Van Hamme} \& {Wilson}(2007)}]{2007ApJ...661.1129V}
{Van Hamme}, W., \& {Wilson}, R.~E. 2007, \apj, 661, 1129

\bibitem[{{von Zeipel}(1924)}]{1924MNRAS..84..665V}
{von Zeipel}, H. 1924, \mnras, 84, 665

\bibitem[{{Welsh} {et~al.}(2010){Welsh}, {Orosz}, {Seager}, {Fortney},
  {Jenkins}, {Rowe}, {Koch}, \& {Borucki}}]{2010ApJ...713L.145W}
{Welsh}, W.~F., {Orosz}, J.~A., {Seager}, S., {Fortney}, J.~J., {Jenkins}, J.,
  {Rowe}, J.~F., {Koch}, D., \& {Borucki}, W.~J. 2010, \apjl, 713, L145

\bibitem[{{Wilson}(1979)}]{1979ApJ...234.1054W}
{Wilson}, R.~E. 1979, \apj, 234, 1054

\bibitem[{{Wilson}(1994)}]{1994PASP..106..921W}
---. 1994, \pasp, 106, 921

\bibitem[{{Wilson} \& {Devinney}(1971)}]{1971ApJ...166..605W}
{Wilson}, R.~E., \& {Devinney}, E.~J. 1971, \apj, 166, 605

\bibitem[{{Wilson} \& {Sofia}(1976)}]{1976ApJ...203..182W}
{Wilson}, R.~E., \& {Sofia}, S. 1976, \apj, 203, 182

\bibitem[{{Winn} {et~al.}(2010){Winn}, {Fabrycky}, {Albrecht}, \&
  {Johnson}}]{2010ApJ...718L.145W}
{Winn}, J.~N., {Fabrycky}, D., {Albrecht}, S., \& {Johnson}, J.~A. 2010, \apjl,
  718, L145

\bibitem[{{Winn} {et~al.}(2009{\natexlab{a}}){Winn}, {Johnson}, {Albrecht},
  {Howard}, {Marcy}, {Crossfield}, \& {Holman}}]{2009ApJ...703L..99W}
{Winn}, J.~N., {Johnson}, J.~A., {Albrecht}, S., {Howard}, A.~W., {Marcy},
  G.~W., {Crossfield}, I.~J., \& {Holman}, M.~J. 2009{\natexlab{a}}, \apjl,
  703, L99

\bibitem[{{Winn} {et~al.}(2009{\natexlab{b}}){Winn}, {Howard}, {Johnson},
  {Marcy}, {Gazak}, {Starkey}, {Ford}, {Col{\'o}n}, {Reyes}, {Nortmann},
  {Dreizler}, {Odewahn}, {Welsh}, {Kadakia}, {Vanderbei}, {Adams}, {Lockhart},
  {Crossfield}, {Valenti}, {Dantowitz}, \& {Carter}}]{2009ApJ...703.2091W}
{Winn}, J.~N., {et~al.} 2009{\natexlab{b}}, \apj, 703, 2091

\end{thebibliography}

\end{document}